\documentclass{jfm}
\usepackage{graphicx}
\usepackage{newtxtext}
\usepackage{newtxmath}
\usepackage{natbib}
\usepackage{hyperref}
\hypersetup{
    colorlinks = true,
    urlcolor   = blue,
    citecolor  = black,
}

\newcommand{\RomanNumeralCaps}[1]
\linenumbers

\title{Velocity field and cavity dynamics in drop impact experiments}

\author{
    V. Lherm\aff{1},\aff{2}\corresp{\email{vlherm@ur.rochester.edu}},
    R. Deguen\aff{3}
}

\affiliation{
    \aff{1}Department of Earth and Environmental Sciences, University of Rochester, 227 Hutchison Hall, Rochester, NY 14627, USA
    \aff{2}Univ. Lyon, ENSL, UCBL, UJM, CNRS, LGL-TPE, F-69007 Lyon, France
    \aff{3}Univ. Grenoble Alpes, Univ. Savoie Mont Blanc, CNRS, IRD, Univ. Gustave Eiffel, ISTerre, 38000 Grenoble, France
}

\begin{document}
\maketitle

\begin{abstract}
Drop impact experiments allow the modelling of a wide variety of natural processes, from raindrop impacts to planetary impact craters. In particular, interpreting the consequences of planetary impacts requires an accurate description of the flow associated with the cratering process. In our experiments, we release a liquid drop above a deep liquid pool to investigate simultaneously the dynamics of the cavity and the velocity field produced around the air–liquid interface. Using particle image velocimetry, we analyse quantitatively the velocity field using a shifted Legendre polynomial decomposition. We show that the velocity field is more complex than considered in previous models, in relation to the non-hemispherical shape of the crater. In particular, the velocity field is dominated by degrees 0 and 1, with contributions from degree 2, and is independent of the Froude and the Weber numbers when these numbers are large enough. We then derive a semi-analytical model based on the Legendre polynomial expansion of an unsteady Bernoulli equation coupled with a kinematic boundary condition at the crater boundary. This model explains the experimental observations and can predict the time evolution of both the velocity field and the shape of the crater, including the initiation of the central jet.
\end{abstract}

%\begin{keywords}
%Authors should not enter keywords on the manuscript, as these must be chosen by the author during the online submission process and will then be added during the typesetting process (see \href{https://www.cambridge.org/core/journals/journal-of-fluid-mechanics/information/list-of-keywords}{Keyword PDF} for the full list).  Other classifications will be added at the same time.
%\end{keywords}

% {\bf MSC Codes }  {\it(Optional)} Please enter your MSC Codes here

\section{Introduction}
\label{sec:introduction}

When a raindrop splashes on the surface of a pond, it takes less than the blink of an eye for a crater to form beneath the surface, throwing a fluid crown into the air, and for it to collapse, propelling upwards a fluid jet. These are the key features of the splashing regime, which occurs within a specific range of drop radius, impact velocity, impact angle, and physical properties of the fluids such as surface tension, density and viscosity \citep{rein_1993}.
\citet{worthington_1908} was the first to report these features using pioneering high-speed photography methods. The splashing regime was then extensively investigated, regarding, in particular, the time evolution of the transient crater following the impact \citep[\textit{e.g.}][]{engel_1967,morton_2000,bisighini_2010}, and the scaling of the maximum crater radius \citep[\textit{e.g.}][]{macklin_1976,engel_1966,lherm_2022}. The formation, evolution and fragmentation of the fluid crown \citep[\textit{e.g.}][]{allen_1975,krechetnikov_2009,zhang_2010,agbaglah_2013} and of the central jet \citep[\textit{e.g.}][]{fedorchenko_2004,ray_2015,vanrijn_2021} have also been examined.

The drop impact processes cover a wide variety of applications. This includes engineering applications such as the water entry of projectiles \citep{clanet_2004} or spray painting \citep{hines_1966}. This also includes Earth sciences applications such as the production of oily marine aerosol by raindrops \citep{murphy_2015}, spray generation from raindrop impacts on seawater and soil \citep{zhou_2020}, and planetary impact craters \citep{melosh_1989,landeau_2021,lherm_2021a,lherm_2022}. Planetary impacts occur on terrestrial planets from the early stages of accretion to modern meteorite impacts. During planetary formation, thermal and chemical partitioning between the core and the mantle is influenced by the physical mechanisms of segregation between the metal of the impactors’ core and the silicates of the growing planet \citep{stevenson_1990,rubie_2015,lherm_2018}, with major implications on the chemical, thermal and magnetic evolution of the planet \citep{fischer_2015,badro_2018,olson_2022}. In particular, the cratering process is responsible for the initial dispersion and mixing of the impactors’ core \citep{landeau_2021,lherm_2022}. In planetary science, impact craters are also a tool to sample the shallow interior of planets and satellites by combining observations of planetary surfaces with excavation and ejecta deposition models \citep{maxwell_1977,barnhart_2011,kurosawa_2019}. Therefore, understanding the implications of these planetary impacts requires to model the velocity field produced during the formation of the crater.

In the splashing regime, the fate of the crater, the fluid crown and the central jet is directly related to the velocity field produced around the crater boundary. The dynamics of the crater is indeed closely related to the velocity field in the ambient fluid, in particular regarding the evolution of the shape of the cavity. The formation of the fluid crown is also related to the ambient velocity field through the mass flux distribution across the initial water surface. Finally, the production of the central jet is associated with a convergent velocity field, resulting from the collapse of the crater due in part to buoyancy forces.

The velocity field associated with the crater evolution in the splashing regime has been investigated both experimentally and numerically in previous studies.
\citet{engel_1962} was the first to examine the velocity field around the crater by seeding the flow with particles in order to visualize the flow streamlines. These observations allowed to determine the velocity field configuration associated with the crater expansion and its subsequent collapse. More recently, the velocity field was investigated using modern Particle Image Velocimetry (PIV) methods. These velocity field measurements have been used to investigate the origin of vortex rings beneath the crater \citep{liow_2009}, the formation of the central jet \citep{vanrijn_2021}, or solutocapillary flows following the impact of drops on salted water \citep{musunuri_2017}. 
Numerical simulations have also focused on the crater velocity field, regarding in particular the entrapment of air bubbles when the crater collapses and the formation of the central jet \citep{morton_2000,ray_2015}. 

Most of the models involving a prediction of the crater velocity field assume either an arbitrary velocity field \citep{maxwell_1977} or an arbitrary velocity potential associated with an imposed crater geometry, such as a hemispherical crater \citep{engel_1967,leng_2001} or a spherical crater able to translate vertically \citep{bisighini_2010}.
Since these models have only been compared with experimental measurements of the crater size and/or shape, a comparison with experimental measurements of the velocity field is thus required to assess their accuracy. In any event, a new model is required to consistently model the geometry of the cavity without the use of an arbitrarily imposed velocity field or potential.

In this paper, we examine simultaneously the dynamics of the cavity and of the velocity field produced in drop impact experiments. In § \ref{sec:experiments}, we present the experimental setup, methods and diagnostics, as well as the set of dimensionless numbers used in this study. In § \ref{sec:results}, we describe the experimental results obtained for the crater shape and the velocity field. In § \ref{sec:models}, we compare the existing velocity field models with our experimental measurements. In § \ref{sec:legendre}, we finally derive a Legendre polynomials model based on an unsteady Bernoulli equation coupled with a kinematic boundary condition.

\section{Experiments}
\label{sec:experiments}

\subsection{Experimental setup}

\begin{figure}
    \centering
    \includegraphics[width=0.7\linewidth]{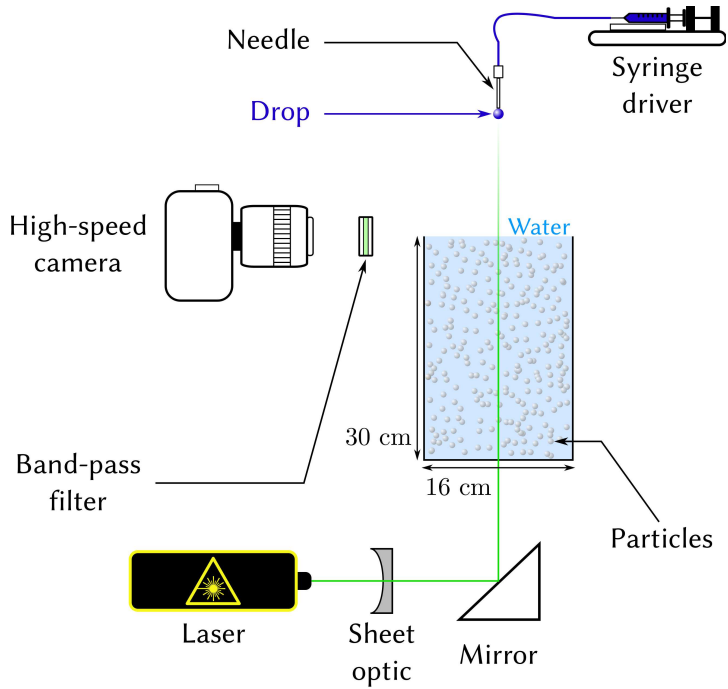}
    \caption{Schematic view of the drop impact experimental setup.}
    \label{fig:Fig1}
\end{figure}

In these experiments, we release a liquid drop in the air above a deep liquid pool of the same liquid (figure \ref{fig:Fig1}). We vary the impact velocity $U_i$ by changing the release height of the drop while keeping the drop radius $R_i$ fixed. We also keep constant the density $\rho$, the viscosity $\mu$ and the surface tension $\sigma$ of the fluids.

The liquid pool is contained in a $16 \times 16 \times 30$ cm glass tank. The pool level is set at the top of the tank to minimise the thickness of the meniscus on the sides of the tank. This allows to image a field of view unperturbed by the free surface meniscus effect.
We generate the drops using a needle supplied with fluid by a syringe driver. When the weight of the drop exceeds the surface tension forces, the drop comes off. We use a nylon plastic needle with an inner diameter of 4.7 mm, generating drops with a radius $R_i=2.7~\mathrm{mm}$. We measured the drop size based on a calibration using mass measurements of dozens of drops and assuming the drop is spherical. We validate this method using high-speed pictures of the drop prior to impact where we can directly measure the drop radius. We obtain a relative difference of 1.4\% between mass measurements and direct measurements.
Impact velocities are in the range $U_i=1-5~\mathrm{m.s^{-1}}$. We calculate the impact velocity for each experiment using a calibrated free-fall model for the drop, including a quadratic drag. We validate this method using high-speed pictures of the drop prior to impact where we can directly measure the drop velocity. We obtain a relative difference of 0.6\% between the velocity model and direct measurements.
We use water both in the drop and in the pool, in a temperature-controlled environment. The density is $\rho=998\pm1~\mathrm{kg.m^{-3}}$. It was measured using an Anton Paar DMA 35 Basic densitometer. The viscosity is $\mu=1\pm0.01~\mathrm{mPa.s}$ \citep{haynes_2016}. The surface tension at the air-water interface is $\sigma=72.8\pm0.4~\mathrm{mJ.m^{-2}}$ \citep{haynes_2016}.

In our experiments, we position the camera at the same height as the water surface. We record images at 1400 Hz with a $2560\times1600$ pixels resolution ($21~\mathrm{\mu m/px}$) and a 12 bits dynamic range, using a high-speed Phantom VEO 640L camera and a Tokina AT-X M100 PRO D Macro lens.

\subsection{Dimensionless numbers}

In these experiments, the impact dynamics depends on $U_i$, $R_i$, $\rho$, $\mu$, $\sigma$, and the acceleration of gravity $g$. Since these six parameters contain three fundamental units, the Vaschy-Buckingham theorem dictates that the impact dynamics depends on a set of three independent dimensionless numbers. We choose the following set:
\begin{equation}
    Fr=\frac{U_i^2}{g R_i}, \quad
    We=\frac{\rho U_i^2R_i}{\sigma}, \quad
    Re=\frac{\rho U_i R_i}{\mu}.
    \label{eq:dimensionless_numbers}
\end{equation}
The Froude number $Fr$ is a measure of the relative importance of impactor inertia and gravity forces. It can also be interpreted as the ratio of the kinetic energy $\rho R_i^3 U_i^2$ of the impactor to its gravitational potential energy $\rho g R_i^4$ just before impact. The Weber number $We$ compares the impactor inertia and interfacial tension at the air-liquid interface. The Reynolds number $Re$ is the ratio between inertial and viscous forces.
In the following, time, lengths and velocities are made dimensionless using the drop radius and the impact velocity, \textit{i.e.} using respectively $R_i/U_i$, $R_i$, $U_i$. These dimensionless quantities are denoted with a tilde. For example, we use a dimensionless time $\tilde{t}=t/(R_i/U_i)$.

We focus on four cases with Froude numbers, Weber numbers and Reynolds numbers respectively in the range $Fr \simeq 100 - 1000$, $We \simeq 100 - 1000$ and $Re \simeq 4400 - 13600$ (table \ref{tab:dimensionless_numbers}). For each case, we conducted three acquisitions, with similar experimental results regarding both the crater shape (\textit{e.g.} figure \ref{fig:Fig4}) and the velocity field (\textit{e.g.} figure \ref{fig:Fig11}). This validates the repeatability of the experiments.

\begin{table}
    \centering
    \begin{tabular}{ccccc}
        Case & A & B & C & D\\
        \hline
        $Fr$ & $103$ & $444$ & $706$ & $979$ \\
        $We$ & $100$ & $429$ & $682$ & $946$ \\
        $Re$ & $4.41 \times 10^3$ & $9.15 \times 10^3$ & $1.15 \times 10^4$ & $1.36 \times 10^4$
    \end{tabular}
    \caption{Dimensionless numbers used in the experiments (see equation \ref{eq:dimensionless_numbers} for details).}
    \label{tab:dimensionless_numbers}
\end{table}

\subsection{Particle Image Velocimetry}

The velocity field is obtained using PIV. We seed the tank with polyamide particles (figure \ref{fig:Fig1}), the concentration, diameter and density of which being respectively $C_p=0.26~\mathrm{g.L^{-1}}$, $d_p=20~\mathrm{\mu m}$ and $\rho_p=1030~\mathrm{kg.m^{-3}}$. We illuminate these particles in suspension with a $1~\mathrm{mm}$ thick laser sheet ($532~\mathrm{nm}$), produced using a continuous $10~\mathrm{W}$ Nd:YAG laser, together with a diverging cylindrical lens and a telescope. The laser sheet is verticalised using a $45^\circ$ inclined mirror located below the tank. The laser wavelength is isolated using a band-pass filter ($532 \pm 10~\mathrm{nm}$).

In order to calculate the velocity field, the camera records two images of the field of view separated by a short time ($\Delta t=200~\mathrm{\mu s}$). These two images are divided into interrogation windows in which a cross-correlation operation allows to obtain the average particle displacement. This involves a five-stage multi-pass processing with interrogation windows decreasing in size. The final interrogation window size is a 64 px square with an overlap of 75\%. In each window, a velocity vector is then calculated, which allows to construct the velocity field over the whole field of view. Finally, the velocity field is spatially calibrated using a sight.

\subsection{Experimental diagnostic}

\subsubsection{Crater shape}
\label{sec:experiments_diagnostic_crater}

\begin{figure}
    \centering
    \includegraphics[width=0.9\linewidth]{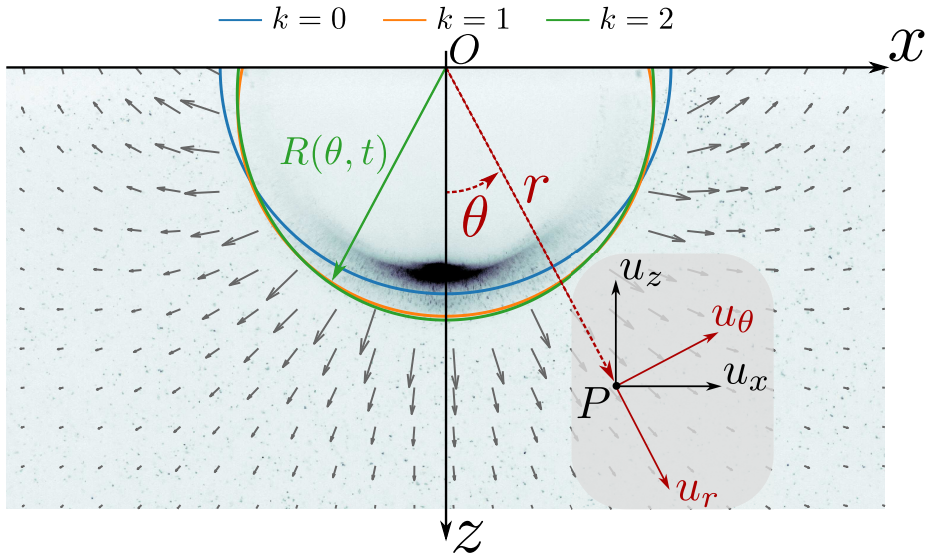}
    \caption{Velocity field resulting from the PIV procedure, superposed on a corresponding experimental raw image.
    The solid lines correspond to the shape of the crater obtained from the shifted Legendre polynomials decomposition (equation \ref{eq:crater_shape}), for degrees $k=0$ (blue), $k=1$ (orange) and $k=2$ (green).
    The definitions of the Cartesian $(x,y,z)$ (black) and of the spherical $(r,\theta,\varphi)$ (red) coordinate systems are also represented.}
    \label{fig:Fig2}
\end{figure}

The crater shape is directly obtained from the raw images used in the PIV procedure (figure \ref{fig:Fig2}). The crater corresponds to a particle-free area, together with a high light intensity area, explained by reflections at the air-water interface, in particular at the bottom of the crater. The crater boundary is defined using these image properties, which allow to delineate the cavity using background removal, an intensity threshold method and image binarisation.

We fit the crater boundary position $R$ (figure \ref{fig:Fig2}), which depends on the polar angle $\theta$ and time $t$, using a set of shifted Legendre polynomials $\bar{P}_k$ up to degree $k_{max}=2$
\begin{equation}
    R(\theta,t)=\sum_{k=0}^{k_{max}} R_k(t)\bar{P}_k(\cos\theta),
    \label{eq:crater_shape}
\end{equation}
where $R_k(t)$ are coefficients fitted with a least-square method. The shifted Legendre polynomials are an affine transformation of the standard Legendre polynomials $\bar{P}_k(x)=P_k(2x-1)$, and are orthogonal on $[0,1]$, \textit{i.e.} on a half-space.
The coefficients $R_k(t)$ correspond to increasingly small scale deviations from a hemispherical shape. $R_0(t)$ corresponds to the mean crater radius (figure \ref{fig:Fig2}, blue line). $R_1(t)$ corresponds to a deformation of the crater, linear in $\cos\theta$, with respect to an hemisphere (figure \ref{fig:Fig2}, orange line). When $R_1(t)>0$, the crater is stretched vertically, resulting in a prolate cavity. When $R_1(t)<0$, the crater is stretched horizontally, resulting in an oblate cavity. Finally, $R_2(t)$ corresponds to a deformation of the crater, quadratic in $\cos\theta$, with respect to a hemisphere (figure \ref{fig:Fig2}, green line).

\begin{figure}
    \centering
    \includegraphics[width=\linewidth]{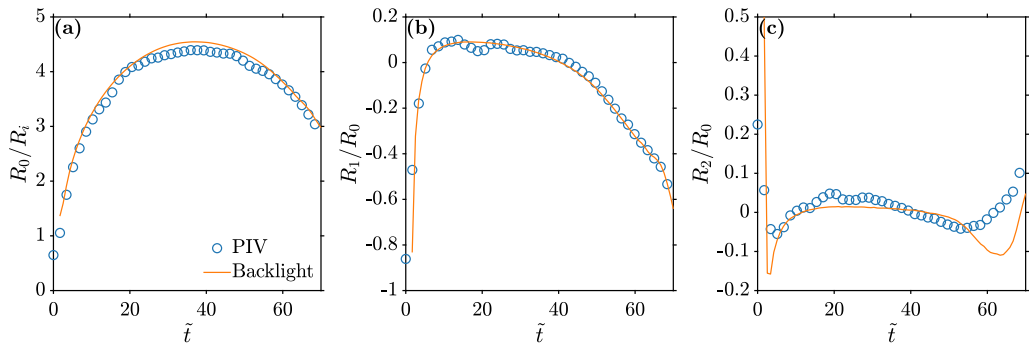}
    \caption{Coefficients $R_0(t)/R_i$ (a), $R_1(t)/R_0$ (b) and $R_2(t)/R_0$ (c) as a function of time $\tilde{t}$. The circles and the solid lines correspond to the crater shape obtained respectively from the PIV procedure (case B, $Fr=444$) and a similar backlight experiment ($Fr=442$).}
    \label{fig:Fig3}
\end{figure}

In order to validate the crater shape determination procedure, we compare the coefficients $R_k(t)$ obtained from the raw images used in the PIV procedure (\textit{e.g.} figure \ref{fig:Fig2}), with the coefficients obtained from an experiment in the same condition, but illuminated from behind (\textit{e.g.} figure \ref{fig:Fig13}). This backlight experiment (see \citet{lherm_2022} for experimental details) allows to determine reliably the shape of the crater.
Figure \ref{fig:Fig3} shows that the coefficients are very similar between the two methods, which validates the crater shape determination procedure from PIV raw images.

\subsubsection{Velocity field}

We aim to compare the experimental velocity field obtained using the PIV procedure with velocity models. For that purpose, the velocity field $\boldsymbol{u}=(u_r,u_\theta,u_\varphi)$ is expressed in a spherical coordinate system ($r,\theta,\varphi$) defined such that $u_r$ and $u_\theta$ are in the plane of the laser sheet (figure \ref{fig:Fig2}, red coordinates). The origin of this coordinate system is the contact point between the impacting drop and the target liquid (figure \ref{fig:Fig2}, point O).

We decompose the components of the velocity field on a shifted Legendre polynomials basis
\begin{equation}
    u_r(r, \theta, t)=\sum_{l=0}^{+\infty} u_{r,l}(r, t) \bar{P}_l(\cos\theta),
    \label{eq:ur_legendre_shifted}
\end{equation}
\begin{equation}
    u_\theta(r, \theta, t)=\sum_{l=0}^{+\infty} u_{\theta,l}(r, t) \bar{P}_l(\cos\theta),
    \label{eq:uth_legendre_shifted}
\end{equation}
where $u_{r,l}(r,t)$ and $u_{\theta,l}(r,t)$ are respectively the decomposition coefficients of $u_r$ and $u_\theta$.
The shifted Legendre polynomials $\bar{P}_l(\cos\theta)$ being orthogonal on half-hemispheres ($\theta \in [0, \pi/2]$), we obtain the $u_{r,l}(r,t)$ and $u_{\theta,l}(r,t)$ coefficients using a least-square inversion of the experimental velocity components over the separate half-hemispheres $\theta \geq 0$ and $\theta < 0$, before averaging the results from the left and right half-hemispheres.
Since the flow is close to axisymmetric (\textit{e.g.} figure \ref{fig:Fig2}), the coefficients obtained by the inversion over each half-hemisphere are very close to each other. Assuming an axisymmetric flow, note that $u_{r,0}(r,t)$ is the average of $u_r$ over the full hemisphere. 

\section{Experimental results}
\label{sec:results}

\subsection{Crater shape}

\begin{figure}
    \centering
    \includegraphics[width=\linewidth]{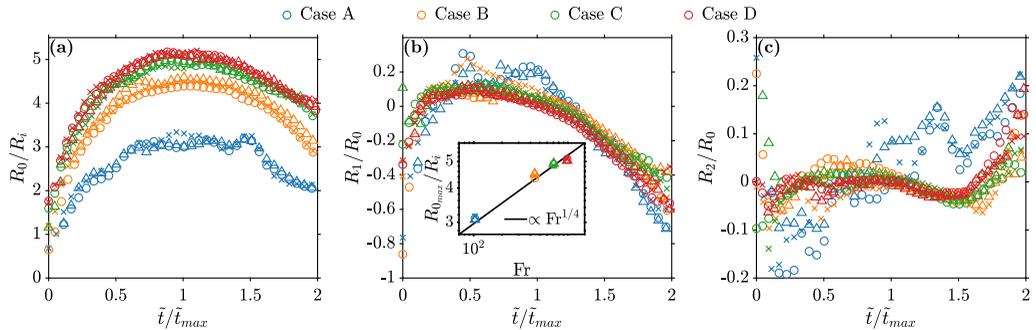}
    \caption{Coefficients $R_0(t)/R_i$ (a), $R_1(t)/R_0(t)$ (b) and $R_2(t)/R_0(t)$ (c) as a function of time normalised by the opening timescale of the crater $\tilde{t}/\tilde{t}_{max}$, in the four cases. Inset: Maximum mean crater radius $R_{0_{max}}/R_i$ as a function of the Froude number $Fr$. For each case, the different types of markers correspond to different experiments.}
    \label{fig:Fig4}
\end{figure}

Figure \ref{fig:Fig4} shows the fitted coefficients of the shifted Legendre decomposition of the crater boundary (equation \ref{eq:crater_shape}) as a function of time, for all experimental cases. We normalise the fitted coefficients $R_1(t)$ and $R_2(t)$ by $R_0(t)$, \textit{i.e.} the mean crater radius. Using this normalisation, we quantify the deviation of the crater geometry from a hemisphere. We also normalise time by the opening timescale of the crater \citep{lherm_2022}
\begin{equation}
    \tilde{t}_{max}=\frac{1}{2}\left(\frac{8}{3}\right)^{1/8}\mathrm{B}\left(\frac{1}{2},\frac{5}{8}\right) \Phi^{1/8} \xi^{1/2} Fr^{5/8},
    \label{eq:energy_model_tmax}
\end{equation}
where $\Phi$ and $\xi$ are respectively energy partitioning and kinetic energy correction coefficients, and $\mathrm{B}$ is the beta function. This scaling is obtained by using an energy conservation equation where the sum of the potential energy of the crater and of the kinetic energy of the crater, corrected by $\xi$, is equal at any instant of time to the kinetic energy of the impacting drop, corrected by $\Phi$. Assuming that the kinetic energy of the crater vanishes when the cavity reaches its maximum size \citep{lherm_2022}, the maximum crater radius scales as
\begin{equation}
    \tilde{R}_{max}=\left(\frac{8}{3}\right)^{1/4}\Phi^{1/4} Fr^{1/4}.
    \label{eq:energy_model_Rmax}
\end{equation}
Using this $Fr^{1/4}$ scaling law,  the energy conservation equation is integrated between $\tilde{t}=0$ and $\tilde{t}=\tilde{t}_{max}$ to obtain the opening timescale of the crater given by equation \ref{eq:energy_model_tmax}. More details can be found in \citet{lherm_2022}.
With our experimental range of Froude number, we use $\Phi=Fr^{-0.156}$ and $\xi=0.34$ \citep{lherm_2022}. This normalisation allows to collapse our experiments on the same timescale.

In figure \ref{fig:Fig4}, the crater shape evolution of case A is markedly different from cases B, C and D. Thus, we describe this case separately.
We first deal with the high $We$ experiments (cases B, C and D), where surface tension effects are negligible in comparison with the impactor inertia \citep[\textit{e.g.}][]{pumphrey_1990,morton_2000,leng_2001,ray_2015}.
The crater size increases with the Froude number (figure \ref{fig:Fig4}, inset), in a way that is compatible with a $Fr^{1/4}$ scaling law for the maximum mean crater radius $R_{0_{max}}$ \citep{engel_1966,leng_2001,lherm_2022}. Furthermore, the evolution of the crater shape relative to the mean crater size is independent of the Froude number, with similar evolution of $R_1(t)/R_0(t)$ and $R_2(t)/R_0(t)$ (figure \ref{fig:Fig4}b-c).

At early times of the crater opening stage ($\tilde{t}/\tilde{t}_{max} \lesssim 0.25$), the mean radius of the crater $R_0(t)$ increases (figure \ref{fig:Fig4}a) as the cavity opens. The crater has a flat-bottomed oblate shape (\textit{e.g.} figure \ref{fig:Fig5}, i) as a result of the spread of the drop on the surface of the pool, with negative $R_1(t)$ (figure \ref{fig:Fig4}b).
The flat-bottomed oblate cavity gradually becomes hemispherical as a result of the overpressure produced at the contact point between the impacting drop and the surface (\textit{e.g.} figure \ref{fig:Fig5}, ii). The magnitude of $R_1(t)/R_0(t)$ indeed decreases with time during this stage (figure \ref{fig:Fig4}b).
The crater is also deformed at higher degrees with mostly negative $R_2(t)/R_0(t)$ (figure \ref{fig:Fig4}c). This corresponds to second-order deviations from the hemispherical shape, with a flattened crater boundary close to the surface (\textit{e.g.} figure \ref{fig:Fig5}, i).

At intermediate times of the crater opening stage ($0.25 \lesssim \tilde{t}/\tilde{t}_{max} \lesssim 0.5$), the crater continues to open (figure \ref{fig:Fig4}a). The cavity is still stretched vertically, which leads to increasingly positive $R_1(t)/R_0(t)$ (figure \ref{fig:Fig4}b), \textit{i.e.} a prolate cavity (figure \ref{fig:Fig5}, iii).
The crater reaches a maximum prolate deformation when $\tilde{t}/\tilde{t}_{max} \simeq 0.5$, with $R_1(t)/R_0(t) \simeq 0.08$ (figure \ref{fig:Fig4}b).
The crater is also deformed at higher degrees, with positive $R_2(t)/R_0(t)$ (figure \ref{fig:Fig4}c). This corresponds to a vertical crater boundary close to the surface (figure \ref{fig:Fig5}, iii).

At late times of the crater opening phase ($0.5 \lesssim \tilde{t}/\tilde{t}_{max} \lesssim 1$), the mean crater radius still increases (figure \ref{fig:Fig4}a) but the crater starts to flatten with decreasing $R_1(t)/R_0(t)$ (figure \ref{fig:Fig4}b). As the opening velocity of the crater decreases, buoyancy forces become significant, resulting in the horizontal stretching of the cavity. The crater flattens to give an approximately hemispherical crater at $\tilde{t}/\tilde{t}_{max} \simeq 1$ (figure \ref{fig:Fig5}, v). 

After the crater has reached its maximum size ($\tilde{t}/\tilde{t}_{max} \gtrsim 1$), the mean crater radius starts to decrease (figure \ref{fig:Fig4}a). $R_1(t)/R_0(t)$ decreases at a rate higher than in the opening stage of the crater (figure \ref{fig:Fig4}b). Horizontal stretching of the crater is accelerated, as expected since buoyancy forces are now prevailing. This leads to the formation of an increasingly oblate cavity (figure \ref{fig:Fig5}, vi-vii).
When $\tilde{t}/\tilde{t}_{max} \gtrsim 1.5$, higher degrees eventually deviate from zero with positive $R_2(t)/R_0(t)$ (figure \ref{fig:Fig4}c). In addition to the negative value of $R_1(t)/R_0(t)$, this corresponds to the formation of the central jet (figure \ref{fig:Fig5}, viii-ix).

We now deal with the moderate $We$ experiment (case A), where surface tension effects are significant in comparison with the impactor inertia \citep[\textit{e.g.}][]{pumphrey_1990,morton_2000,leng_2001,ray_2015}.
In this case, a downward propagating capillary wave develops at the cavity interface and drives the crater deformation, often leading to the entrapment of a bubble due to the pinching of the cavity \citep[\textit{e.g.}][]{oguz_1990,pumphrey_1990,prosperetti_1993,elmore_2001}. This mechanism is typically expected at moderate $We$, \textit{i.e.} $We \simeq 30-140$ \citep[figure 6 in][]{pumphrey_1990}.
During crater opening, this explains why the maximum prolate deformation occurs later than in the other cases, at $\tilde{t}/\tilde{t}_{max} \simeq 0.8$, and why the prolate deformation is larger, with $R_1(t)/R_0(t) \simeq 0.2$ (figure \ref{fig:Fig4}b). During crater closing, the evolution of $R_1(t)/R_0(t)$ and $R_2(t)/R_0(t)$ is markedly different from the other cases due to the convergence of the capillary wave at the bottom of the crater. 

\subsection{Velocity field}

\subsubsection{Velocity maps}

\begin{figure}
    \centering
    \includegraphics[width=\linewidth]{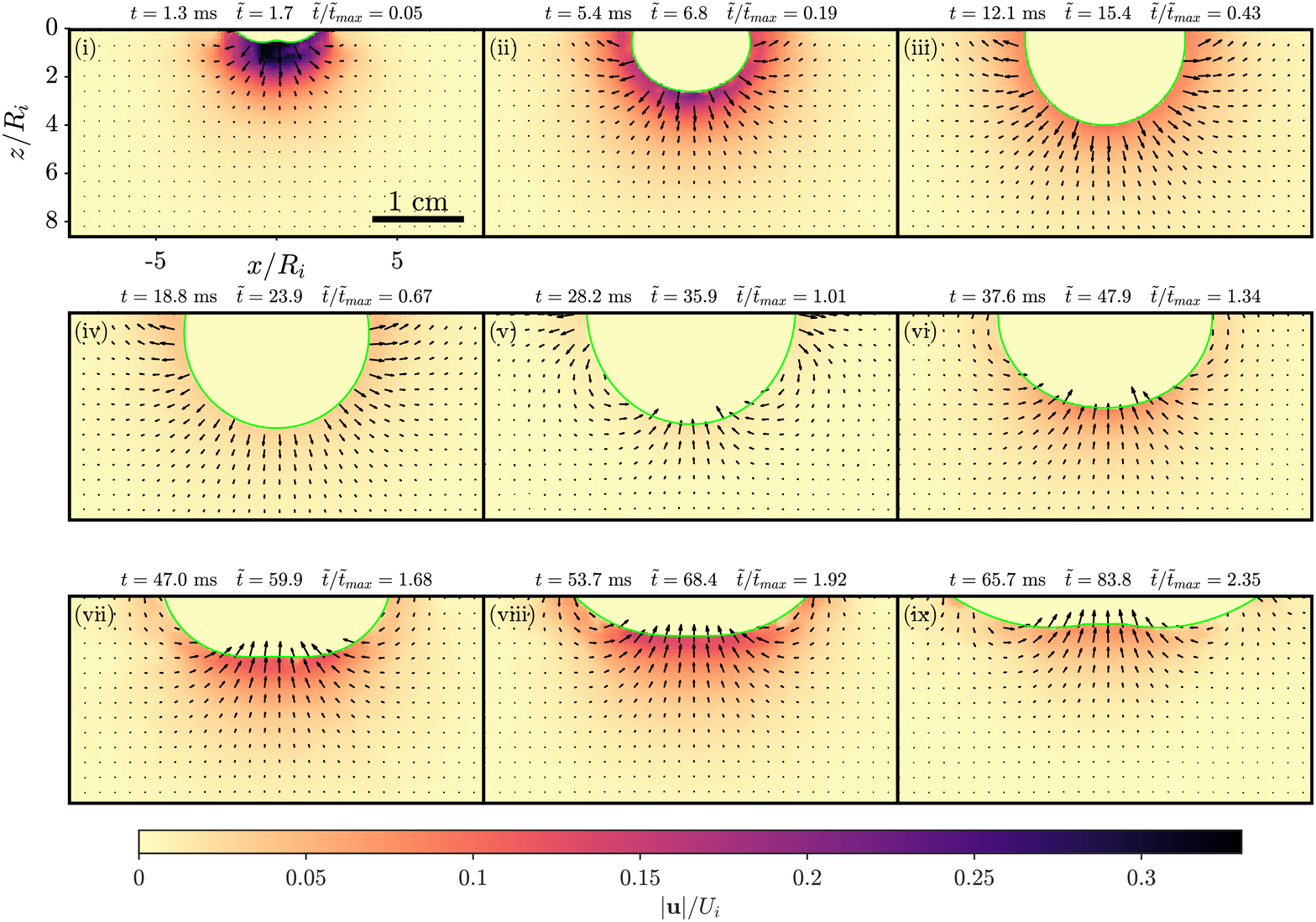}
    \caption{Time evolution of the velocity $|\boldsymbol{u}|=(u_x^2+u_z^2)^{1/2}$ for case B. The vector field corresponds to the experimental velocity field, normalised by its maximum value in each snapshot. The solid green line corresponds to the crater boundary determined using the Legendre polynomial decomposition (equation \ref{eq:crater_shape}).}
    \label{fig:Fig5}
\end{figure}

Figure \ref{fig:Fig5} shows the evolution of the norm of the velocity $|\boldsymbol{u}|=(u_x^2+u_z^2)^{1/2}$ as a function of time, for case B. 
During the opening stage of the crater, the velocity around the crater gradually decreases due to the deceleration of the crater boundary (figure \ref{fig:Fig5}, i-iv). The maximum velocity is $\simeq 1.1~\mathrm{m.s^{-1}}$ at time 1.3 ms after contact (figure \ref{fig:Fig5}, i), which corresponds to 32\% of the impact velocity.
When $\tilde{t}/\tilde{t}_{max} \gtrsim 0.1$, the norm of the velocity decreases radially around the crater (figure \ref{fig:Fig5}, ii-iv), whereas, when $\tilde{t}/\tilde{t}_{max} \lesssim 0.1$, the velocity decreases at a higher rate on the side of the crater. This may be explained by the initial oblate shape of the crater, related to the spread of the drop on the water surface upon impact, which leads to a higher velocity beneath the crater as it becomes gradually hemispherical.
The velocity field is composed of a dominant radial component and of a polar component responsible for an upward flow across the initial water surface (figure \ref{fig:Fig5}, i-iv). The polar component is thus responsible for the formation of the liquid crown above the water surface \citep[\textit{e.g.}][]{rein_1993,fedorchenko_2004,zhang_2010}.

When the crater reaches its maximum size (figure \ref{fig:Fig5}, v), the cavity is nearly hemispherical and the velocity field seems to vanish simultaneously in the entire flow, consistently with \citet{engel_1966}'s observations, which were subsequently used in several velocity models \citep[\textit{e.g.}][]{engel_1967,prosperetti_1993}. However, this first-order assumption on the simultaneous vanishing velocity field does not hold when the flow is examined in detail.
Beneath the cavity, the velocity gradually decreases and eventually vanishes just before the crater reaches its maximum size. The velocity is directed downwards due to the expansion of the crater. The velocity then increases again but is directed upwards due to the collapse of the crater.
On the side of the cavity, close to the surface, the velocity does not vanish when the crater reaches its maximum size. The collapse of the crater takes over its initial expansion, which allows to keep outward velocities on the side of the crater.

When the crater collapses (figure \ref{fig:Fig5}, vi-ix), a convergent flow forms towards the centre of the cavity. This leads to the formation of the central jet.

\begin{figure}
    \centering
    \includegraphics[width=\linewidth]{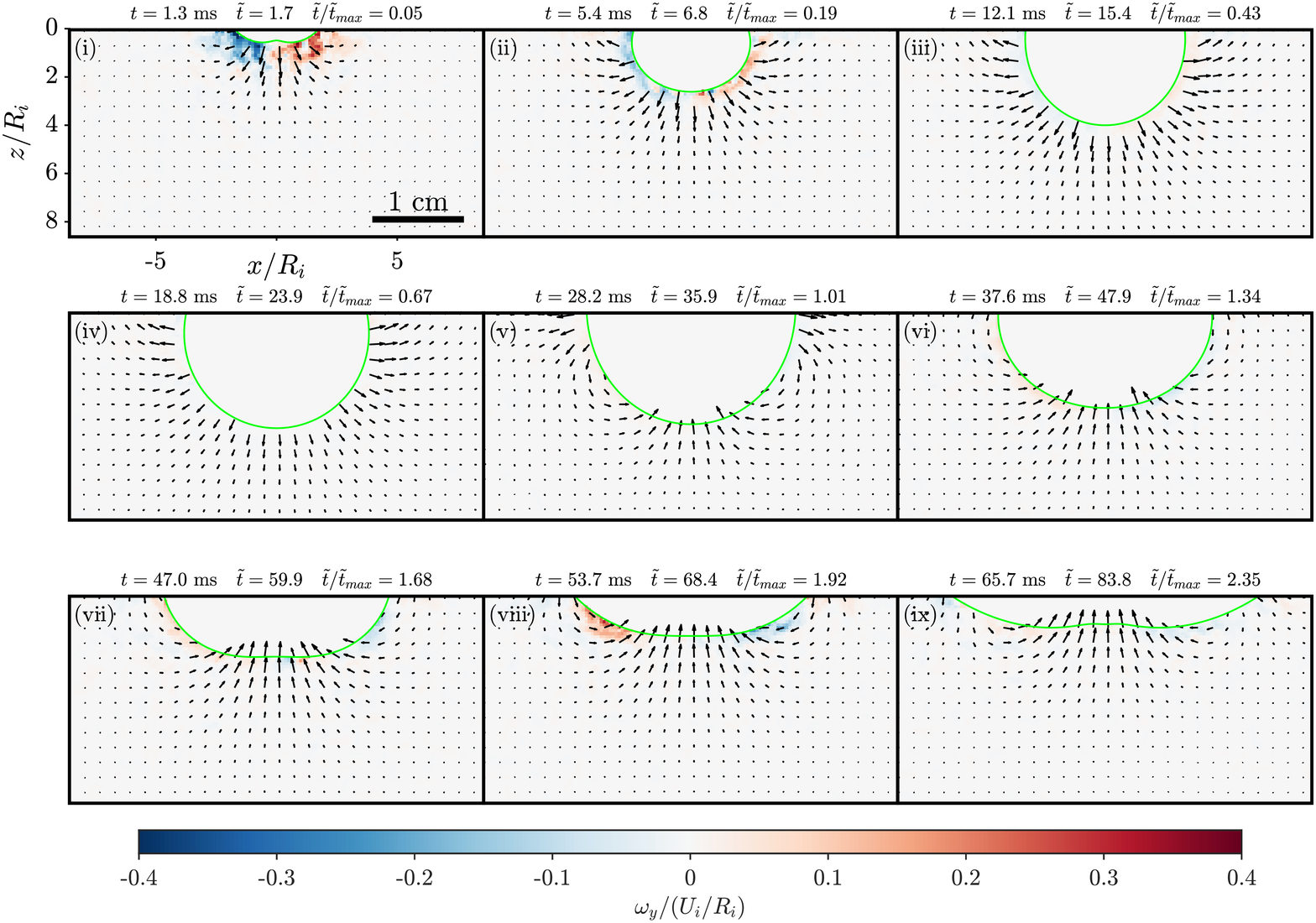}
    \caption{Time evolution of the vorticity $\omega_y=\partial u_x/\partial z-\partial u_z/\partial x$ for case B. The vector field corresponds to the experimental velocity field, normalised by its maximum value in each snapshot. The solid green line corresponds to the crater boundary determined using the Legendre polynomial decomposition (equation \ref{eq:crater_shape}).}
    \label{fig:Fig6}
\end{figure}

Figure \ref{fig:Fig6} shows the evolution of the vorticity $\omega_y=\partial u_x/\partial z-\partial u_z/\partial x$ as a function of time, for case B. 
The vorticity produced by the impact around the crater is confined close to the air-water boundary, in particular when the crater is strongly deformed, at the beginning of the crater opening (figure \ref{fig:Fig6}, i-ii) and when it collapses (figure \ref{fig:Fig6}, vi-ix). This suggests that the flow is mostly irrotational, which supports the potential flow assumption used in previous models (§\ref{sec:models}).
Furthermore, some of the vorticity observed around the crater boundary may be an artefact related to spurious velocity measurements produced by cross-correlations on reflections at the air-water interface, and not on PIV particles. This assumption is supported by the estimated diffusion length of the vorticity (0.3 mm in 100 ms) which is significantly smaller than the typical size of the vorticity band.

\subsubsection{Velocity coefficients}

\begin{figure}
    \centering
    \includegraphics[width=\linewidth]{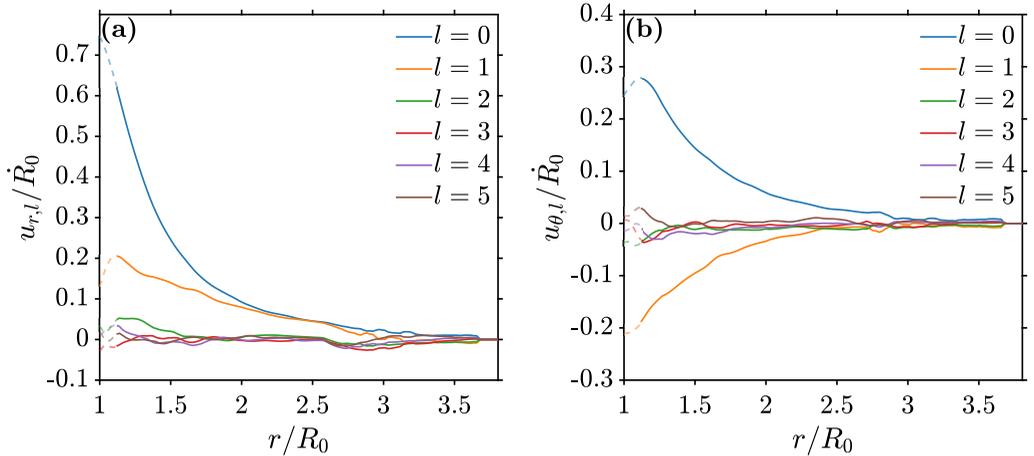}
    \caption{Coefficients $u_{r,l}(r,t)$ and $u_{\theta,l}(r,t)$ normalised by the mean crater velocity $\dot{R}_0(t)$, as a function of the radial coordinate $r$, normalised by the mean crater radius $R_0(t)$, up to degree $l=5$. Dashed lines correspond to regions where the coefficients sample velocity artefacts are located in the crater. The coefficients are calculated at $\tilde{t}=15.4$ ($\tilde{t}/\tilde{t}_{max}=0.43$) for case B.}
    \label{fig:Fig7}
\end{figure}

Figure \ref{fig:Fig7} shows the coefficients $u_{r,l}(r,t)$ and $u_{\theta,l}(r,t)$ (equations \ref{eq:ur_legendre_shifted}-\ref{eq:uth_legendre_shifted}) as a function of the radial coordinate at a given time $\tilde{t}=15.4$ ($\tilde{t}/\tilde{t}_{max}=0.43$) during the crater opening stage of case B. During this stage, the velocity field is dominated by the degrees $l=0$ and $l=1$, the higher degrees $l\geq2$ being much smaller.
When $r \lesssim 1.2R_0$, we observe a decrease in the slope of the coefficients. This may be related to the deviation of the crater from a hemisphere. The coefficients indeed sample points located at varying distances from the actual crater boundary, including artefacts located into the crater, which may influence the radial dependency of these coefficients close to the crater boundary. In figure \ref{fig:Fig7}, we identify this misleading trend by using dashed lines when the radius is smaller than $\max\{R(\theta)\}$ ($r \leq 1.11 R_0$ in figure \ref{fig:Fig7}).

\begin{figure}
    \centering
    \includegraphics[width=\linewidth]{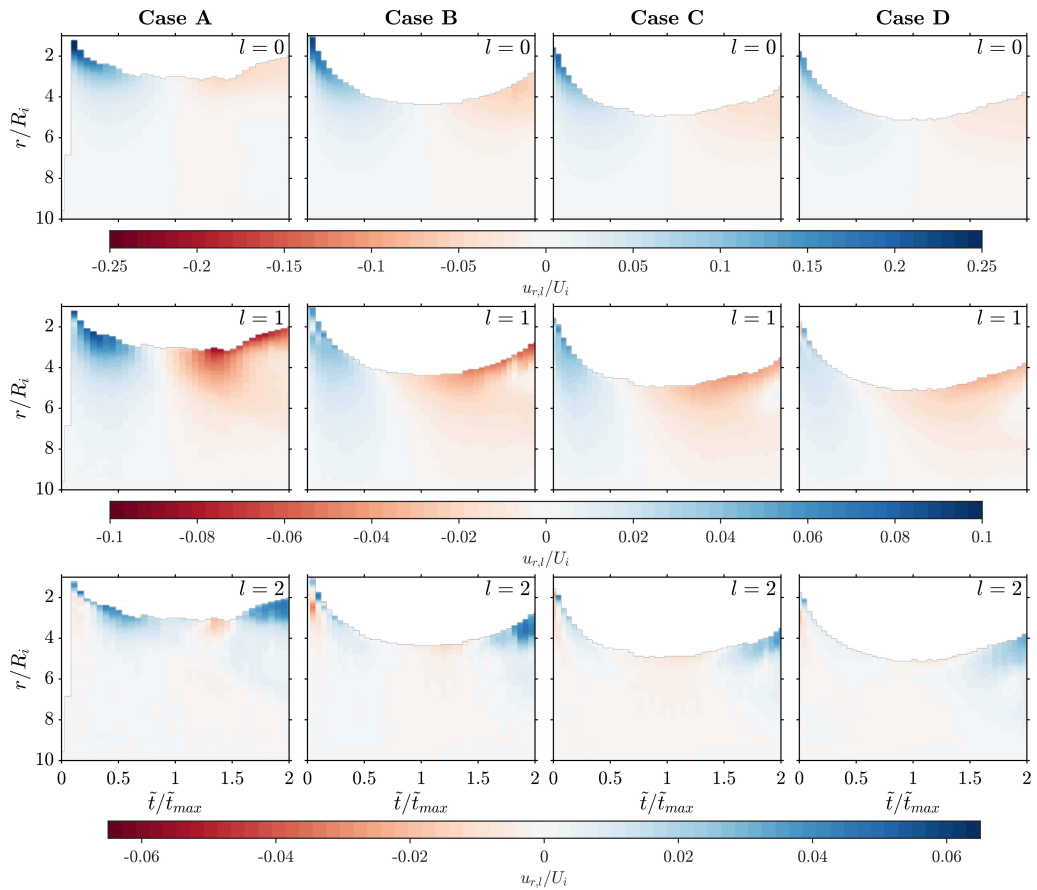}
    \caption{Time evolution of the coefficient $u_{r,l}(r,t)$ ($l\in\{0,1,2\}$) normalised by the impact velocity $U_i$, as a function of the radial coordinate $r$ normalised by the drop radius $R_i$, for case B. Time is normalised by the opening timescale of the crater $\tilde{t}_{max}$.}
    \label{fig:Fig8}
\end{figure}

\begin{figure}
    \centering
    \includegraphics[width=\linewidth]{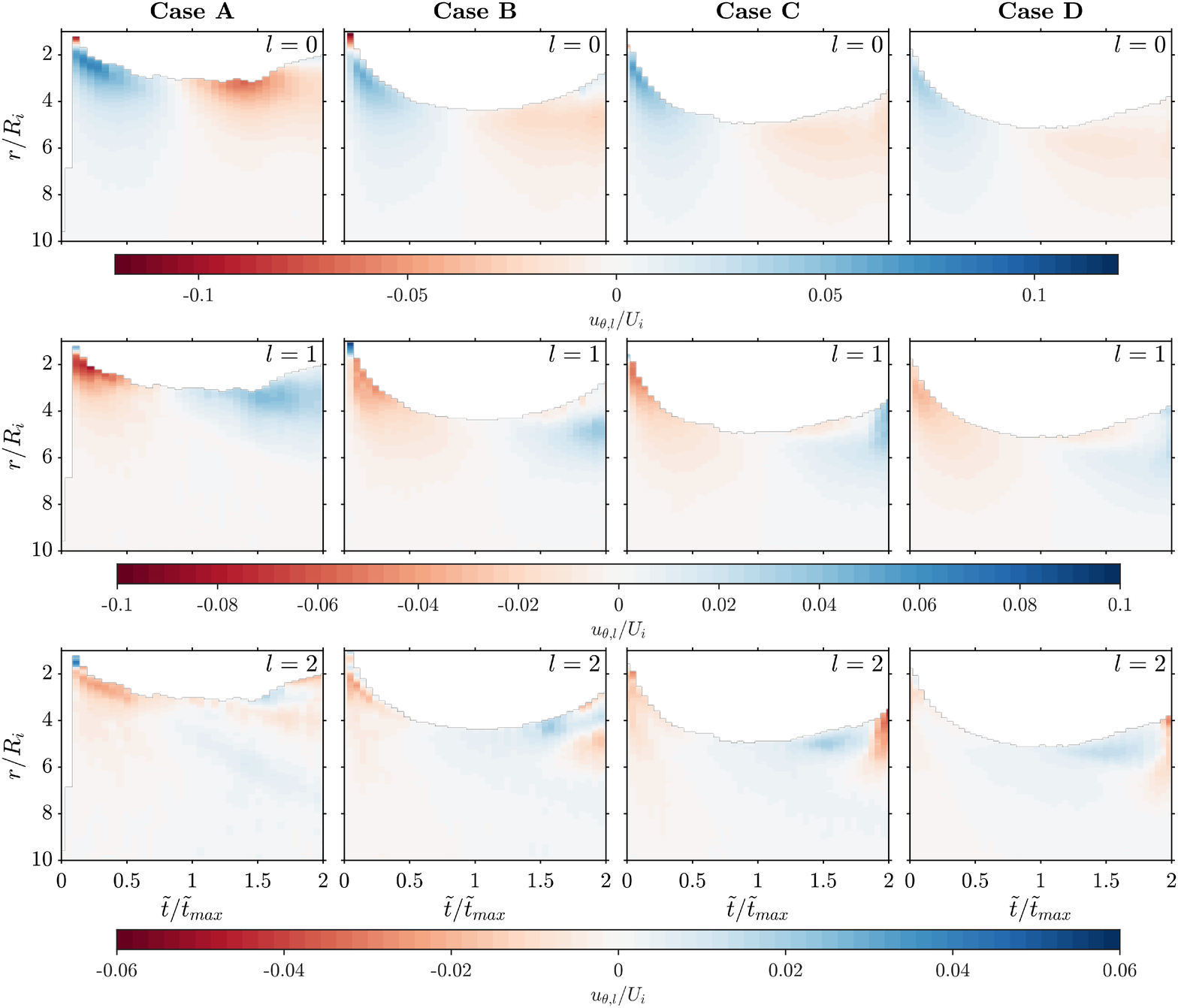}
    \caption{Time evolution of the coefficient $u_{\theta,l}(r,t)$ ($l\in\{0,1,2\}$) normalised by the impact velocity $U_i$, as a function of the radial coordinate $r$ normalised by the drop radius $R_i$, for case B. Time is normalised by the opening timescale of the crater $\tilde{t}_{max}$.}
    \label{fig:Fig9}
\end{figure}

Figures \ref{fig:Fig8} and \ref{fig:Fig9} compare the time evolution of the coefficients $u_{r,l}(r,t)$ and $u_{\theta,l}(r,t)$ (equations \ref{eq:ur_legendre_shifted}-\ref{eq:uth_legendre_shifted}) between the cases, for $l \leq 2$.
Except for the different normalisation, figure \ref{fig:Fig7} is thus similar to a radial slice of these coefficients maps, for case B, at $\tilde{t}/\tilde{t}_{max}=0.43$.
As for the crater shape, the moderate $We$ case A is different from the high $We$ cases B, C and D, both for the radial (figure \ref{fig:Fig8}) and the polar (figure \ref{fig:Fig9}) component of the velocity field. We thus deal with this case separately.

We first deal with the high $We$ experiments (cases B, C and D), where both components of the velocity field are similar among cases, regardless of the degree in question. The velocity field is mostly dominated by the degrees $l=0$ and $l=1$, both during the opening and the closing stage of the crater, in agreement with figure \ref{fig:Fig7}.

During the crater opening stage ($\tilde{t}/\tilde{t}_{max} \lesssim 1$), the dominant degrees of the radial component $u_{r,0}(r,t)$ and $u_{r,1}(r,t)$ are positive (figure \ref{fig:Fig8}). This corresponds to the strong radial velocity field related to the expansion of the cavity.
The dominant degrees of the polar component $u_{\theta,0}(r,t)$ and $u_{\theta,1}(r,t)$ are concomitantly positive and negative, respectively (figure \ref{fig:Fig9}), with a lower magnitude. This corresponds to a polar perturbation of the dominant radial velocity field, related to the mass flux across the surface $z=0$ which produces the fluid crown. The positive coefficient $u_{\theta,0}(r,t)$ indeed corresponds to a flow toward the surface, while the negative coefficient $u_{\theta,1}(r,t)$ corresponds to a degree $l=1$ perturbation, linear in $\cos\theta$.
The degree $l=2$ also contributes to the velocity field of both components, in particular when the crater is strongly deformed due to the spread of the drop at the surface of the pool, at the beginning of the opening stage ($\tilde{t}/\tilde{t}_{max} \lesssim 0.25$).

When the crater reaches its maximum size ($\tilde{t}/\tilde{t}_{max} \simeq 1$), the dominant degrees of both components change signs as the crater starts to collapse. In detail, $u_{r,0}(r,t)$ vanishes later ($\tilde{t}/\tilde{t}_{max} \simeq 1$) than $u_{r,1}(r,t)$ ($\tilde{t}/\tilde{t}_{max} \simeq 0.6$) (figure \ref{fig:Fig8}) and $u_{\theta,0}(r,t)$ ($\tilde{t}/\tilde{t}_{max} \simeq 0.8$) (figure \ref{fig:Fig9}). This is in agreement with the observations of figure \ref{fig:Fig5} at $\tilde{t}/\tilde{t}_{max} \simeq 1$, where the velocity vanishes beneath the crater but not on the sides.

During the crater closing stage ($\tilde{t}/\tilde{t}_{max} \gtrsim 1$), $u_{r,0}(r,t)$ and $u_{r,1}(r,t)$ are both negative (figure \ref{fig:Fig8}) and $u_{\theta,0}(r,t)$ is negative (figure \ref{fig:Fig9}). This corresponds to the development of the convergent flow related to the collapse of the crater and the formation of the central jet.
As at the beginning of the opening stage, the degree $l=2$ of both components contributes significantly to the velocity field at the end of the closing stage ($\tilde{t}/\tilde{t}_{max} \gtrsim 1.5$), in relation with the strongly deformed crater boundary.

We now deal with the moderate $We$ experiment (case A). Although the degree $l=0$ of both components is similar to the high $We$ cases, the degrees $l=1$ and $l=2$ of case A are significantly larger than their counterparts of cases B, C and D. Furthermore, the time at which $u_{r,1}(r,t)$ and $u_{\theta,0}(r,t)$ vanish is significantly modified.
This may also be a consequence of significant surface tension effects in this moderate $We$ experiment, related to vigorous deformations of the crater boundary by the propagation of a capillary wave towards the bottom of the crater.

\section{Comparison with existing velocity models}
\label{sec:models}

\begin{figure}
    \centering
    \includegraphics[width=\linewidth]{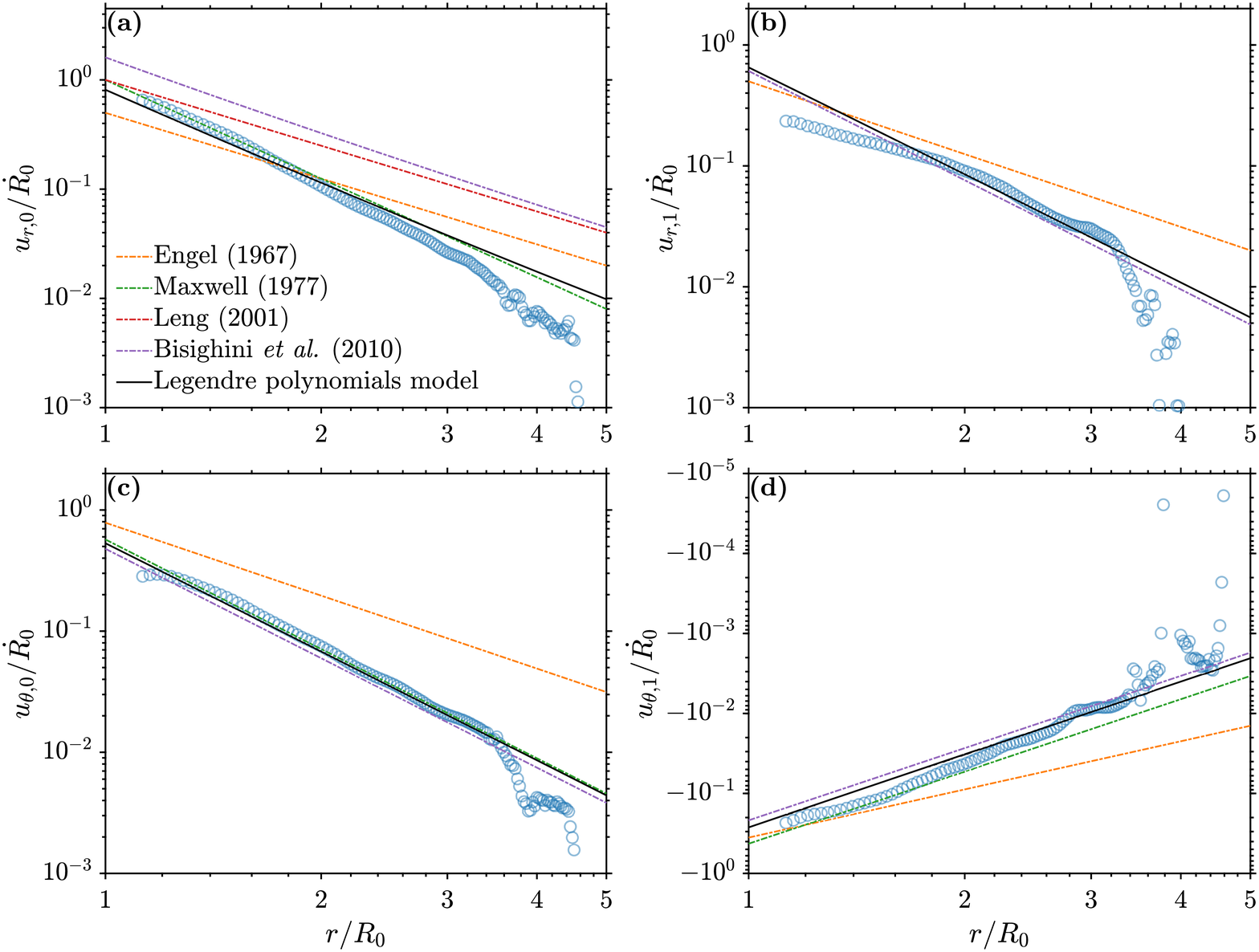}
    \caption{Coefficients $u_{r,0}(r,t)$ (a), $u_{r,1}(r,t)$ (b), $u_{\theta,0}(r,t)$ (c) and $u_{\theta,1}(r,t)$ (d) normalised by the mean crater velocity $\dot{R}_0(t)$, as a function of the radial coordinate $r$, normalised by the mean crater radius $R_0(t)$. The circles correspond to case B at $\tilde{t}/\tilde{t}_{max}=0.24$. The dash-dotted lines correspond to the models of \citet{engel_1967}, \citet{maxwell_1977}, \citet{leng_2001} and \citet{bisighini_2010}. The solid line corresponds to the solution of the predictive model, using the simplified equation system (equation \ref{eq:bernoulli_BC_simplified}) with the reference set of initial conditions (equation \ref{eq:reference_initial_conditions}).}
    \label{fig:Fig10}
\end{figure}

In this section, we review the velocity models proposed by \citet{engel_1967}, \citet{maxwell_1977}, \citet{leng_2001} and \citet{bisighini_2010}, and compare their predictions with our observations. Since most of these models have been designed to understand the crater opening stage, we compare these models with our experimental velocity measurements by focusing on a typical snapshot of this initial stage.
For that purpose, figure \ref{fig:Fig10} shows the dominant coefficients $u_{r,0}(r,t)$, $u_{r,1}(r,t)$, $u_{\theta,0}(r,t)$ and $u_{\theta,1}(r,t)$ of case B as a function of the radial coordinate at $\tilde{t}/\tilde{t}_{max}=0.24$, as well as the predictions of the models.

\subsection{Engel (1967)'s model}

\citet{engel_1967}'s model assumes an energy balance where the potential energy of the crater, the potential energy of a cylindrical wave developing above the surface, the surface tension energy of the produced interface, the kinetic energy of the flow around the crater, the kinetic energy of the cylindrical wave and viscous dissipation are equal at any time to half of the kinetic energy of the impacting drop. Among the assumptions of such a model, \citet{engel_1967} assumes a hemispherical crater with a radius $R_0(t)$ and a potential flow with a velocity potential $\phi$ satisfying the boundary conditions on the velocity $|\boldsymbol{u}|(r=+\infty)=0$ and $|\boldsymbol{u}|[r=R_0(t)]=\dot{R}_0(t)$. The velocity potential used in the model is
\begin{equation}
    \phi=-\frac{\dot{R}_0R_0^2\cos\theta}{r}.
    \label{eq:potential_engel}
\end{equation}
The radial component $u_r$ and the polar component $u_\theta$ of the velocity field, obtained by deriving the velocity potential, write
\begin{equation}
    \left\{
    \begin{array}{l}
        u_r=\frac{\dot{R}_0R_0^2\cos\theta}{r^2}\\
        u_\theta=\frac{\dot{R}_0R_0^2\sin\theta}{r^2}
    \end{array}
    \right..
    \label{eq:velocity_engel}
\end{equation}

This model allows to capture the evolution of the mean crater radius \citep[\textit{e.g.}][figure 3]{engel_1967}. The velocity field has a $l=0$ (figure \ref{fig:Fig10}a) and $l=1$ (figure \ref{fig:Fig10}b) radial components and a $l=0$ (figure \ref{fig:Fig10}c) and $l=1$ (figure \ref{fig:Fig10}d) polar components. This allows to obtain a velocity field qualitatively similar to the experiments, including in particular a degree $l=1$ of the radial component, and a polar component. However, the slopes of the velocity components are smaller than the experimental slopes, in particular the $1/r^2$ slope of $u_{r,0}(r,t)$.
The main limitations of \citet{engel_1967}'s model are the fixed hemispherical geometry of the crater and the arbitrary velocity potential defined to fit experimental observations of the velocity field. 
More importantly, this velocity potential (equation \ref{eq:potential_engel}) corresponds to the flow around an expanding cylinder (with $r$ being the distance from the cylinder axis, and $\theta$ the angular position around this axis) rather than around an expanding sphere, as incorrectly assumed in \citet{engel_1967}. It is not a solution of the Laplace equation in spherical coordinates and has a non-zero divergence.

\subsection{Maxwell (1977)'s model}

\citet{maxwell_1977}'s model assumes an empirical form of the velocity field based on planetary cratering observations. The model assumes that the radial component $u_r$ is independent of $\theta$ and that its radial dependency is a power $Z$ of the radius $r$. $u_\theta$ is then calculated using fluid incompressibility. The velocity field thus writes
\begin{equation}
    \left\{
    \begin{array}{l}
        u_r=\frac{\alpha(t)}{r^Z}\\
        u_\theta=(Z-2)\frac{\sin\theta}{1+\cos\theta}\frac{\alpha(t)}{r^Z}
    \end{array}
    \right.,
    \label{eq:velocity_maxwell}
\end{equation}
where $\alpha(t)$ is an arbitrary coefficient corresponding to the time-dependent flow intensity. According to \citet{maxwell_1977} and \citet{melosh_1989}, the value $Z=3$ gives a velocity field consistent with numerical simulations of explosion and planetary impacts.

This model allows to predict the experimental $u_{r,0}(r,t)$ (figure \ref{fig:Fig10}a), $u_{\theta,0}(r,t)$ (figure \ref{fig:Fig10}c) and $u_{\theta,1}(r,t)$ (figure \ref{fig:Fig10}d) using $Z=3$ and $\alpha(t)=1$. In particular, the slopes predicted by the model are very close to the experimental slopes. However, this model does not allow a degree $l=1$ of the radial velocity component.
The main limitations of \citet{maxwell_1977}'s model are the arbitrary choice for the model time-dependency, with $\alpha(t)$, the fact that $Z$ could depend on $\theta$, which would yield a degree $l=1$ for $u_r$, and the fact that Maxwell's flow is not potential, which is inconsistent with the experimental results (figure \ref{fig:Fig6}).

\subsection{Leng (2001)'s model}

\citet{leng_2001}'s model is similar to  \citet{engel_1967}'s model since it uses a hemispherical crater with a radius $R_0(t)$ and a potential flow. The velocity potential $\phi$ writes
\begin{equation}
    \phi=-\frac{\dot{R}_0R_0^2}{r},
    \label{eq:potential_leng}
\end{equation}
which allows to obtain the velocity components $u_r$ and $u_\theta$ of the velocity field
\begin{equation}
    \left\{
    \begin{array}{l}
        u_r=\frac{\dot{R}_0R_0^2}{r^2}\\
        u_\theta=0
    \end{array}
    \right..
    \label{eq:velocity_leng}
\end{equation}
This velocity potential satisfies the boundary conditions and is a solution of the Laplace equation in spherical coordinates.

This model allows, in particular, to capture the evolution of the mean crater radius using an energy balance, although it requires to multiply the kinetic energy and the total energy by empirical correction factors \citep[\textit{e.g.}][]{lherm_2022}. However, the velocity field has only a degree $l=0$ (figure \ref{fig:Fig10}a) on the radial component and no polar component. As for \citet{engel_1967}'s model, the $1/r^2$ slope of $u_{r,0}(r,t)$ is smaller than the experimental slope. The main limitations of \citet{leng_2001}'s model are the hemispherical geometry and the oversimplified velocity potential which prevents a polar dependency of the radial component and a polar component of the velocity field.

\subsection{Bisighini et al. (2010)'s model}

\citet{bisighini_2010}'s model assumes an expanding spherical crater able to translate vertically over time, with a radius $R_0(t)$ and a vertical position of the crater barycenter $z_c(t)$. This allows to define a velocity potential $\phi$ which corresponds to the superposition between the radial expansion of the crater and the flow past a translating sphere. This potential satisfies the boundary conditions and the Laplace equation in spherical coordinates. In the moving sphere coordinate system $(r',\theta')$, it writes
\begin{equation}
    \phi=-\frac{\dot{R}_0R_0^2}{r'}-\dot{z}_c r'\left(1-\frac{R_0^3}{2r'^3}\right)\cos\theta',
    \label{eq:potential_bisighini}
\end{equation}
with components $u_r$ and $u_\theta$ of the velocity field writing
\begin{equation}
    \left\{
    \begin{array}{l}
        u_r=\frac{\dot{R}_0R_0^2}{r'^2}-\left(1-\frac{R_0^3}{r'^3}\right)\dot{z}_c\cos\theta'\\
        u_\theta=\left(1+\frac{R_0^3}{2r'^3}\right)\dot{z}_c\sin\theta'
    \end{array}
    \right..
    \label{eq:velocity_bisighini}
\end{equation}
\citet{bisighini_2010} then use an unsteady Bernoulli equation to determine the evolution of the sphere radius and position over time. To compare \citet{bisighini_2010}'s model with our experimental data, we need to calculate the corresponding velocity field in the fixed frame of reference by adding the velocity of the crater barycenter $\dot{z}_c (\cos\theta, -\sin\theta)$ to equation \ref{eq:velocity_bisighini}, and expressing $r'$ and $\theta'$ as functions of $r$ and $\theta$ ($r'=\sqrt{r^2+z_c^2-2 z_c r \cos\theta}$, $\cos\theta'=(r\cos\theta - z_c)/r'$, $\sin\theta'=r\sin\theta/r'$).

The velocity field has a $l=0$ (figure \ref{fig:Fig10}a) and a $l=1$ (figure \ref{fig:Fig10}b) radial component, as well as a $l=0$ (figure \ref{fig:Fig10}c) and a $l=1$ (figure \ref{fig:Fig10}d) polar component. The coefficients are calculated using $z_c=0$ and $\dot{z}_c=0.2U_i$, which corresponds to typical values during crater opening (\textit{e.g.} figure \ref{fig:Fig5}). This model explains relatively well the shape of the crater \citep[\textit{e.g.}][figure 17]{bisighini_2010}, and the key tendencies of the experimental components of the velocity field.
However, \citet{bisighini_2010}'s model strongly constrains the geometry of the crater, as well as the related velocity potential definition. As in \citet{engel_1967}'s and \citet{leng_2001}'s models, the $1/r^2$ slope of $u_{r,0}$ is smaller than the experimental slope.

\subsection{Towards a new model}

In all models, either the geometry of the velocity field \citep{engel_1967,maxwell_1977,leng_2001} or the shape of the cavity \citep{engel_1967,leng_2001,bisighini_2010} are imposed. This leads in particular to an incorrect radial dependency of $u_r$, with an exponent much larger in the experiments than in the models, except for \citet{maxwell_1977}'s model where the radial dependency is arbitrarily imposed by the parameter $Z$.
The experimental observation that the radial velocity field decreases with $r$ faster than $1/r^2$ is unexpected since it suggests that the flow component associated with an isotropic expansion of the cavity ($\propto 1/r^2$) is not dominant.
New models are thus required to explain the geometry of the experimental velocity field, as well as the evolution of the non-hemispherical shape of the cavity. In the following section, we develop a semi-analytical model based on a Legendre polynomials expansion of an unsteady Bernoulli equation, coupled with a kinematic boundary condition at the crater boundary.

\section{Legendre polynomials model}
\label{sec:legendre}

In this model, we assume that the fluid is inviscid (\textit{i.e.} $\mu=0$), incompressible (\textit{i.e.} $\nabla\cdot\boldsymbol{u}=0$), and that the flow is irrotational (\textit{i.e.} $\nabla\times\boldsymbol{u}=0$). This means that the flow is potential and satisfies the Laplace equation $\nabla^2\phi=0$, where $\phi$ is the velocity potential defined as $\boldsymbol{u}=\nabla\phi$. In the spherical coordinate system $(r,\theta,\varphi)$, assuming an axisymmetric flow, the solution of the Laplace equation writes
\begin{equation}
    \phi(r,\theta,t)=\sum_{n=0}^{+\infty}\frac{\phi_n(t)}{r^{n+1}}P_n(\cos\theta),
    \label{eq:potential_legendre}
\end{equation}
where $\phi_n(t)$ are time-dependent coefficients and $P_n(x)$ are the standard Legendre polynomials, orthogonal on $[-1,1]$. The components $u_r$ and $u_\theta$ of the velocity field then writes
\begin{equation}
    \left\{
    \begin{array}{rcccl}
        u_r(r,\theta,t)&=&\frac{\partial \phi}{\partial r}&=&
        \sum_{n=0}^{+\infty}-(n+1)\frac{\phi_n(t)}{r^{n+2}}P_n(\cos\theta) \\
        u_\theta(r,\theta,t)&=&\frac{1}{r}\frac{\partial \phi}{\partial\theta}&=&
        \sum_{n=0}^{+\infty}\frac{\phi_n(t)}{r^{n+2}}\frac{\partial P_n(\cos\theta)}{\partial\theta}
    \end{array}
    \right..
    \label{eq:velocity_legendre}
\end{equation}
We also assume a non-hemispherical crater, where the shape of the cavity is decomposed on a set of shifted Legendre polynomials (equation \ref{eq:crater_shape}).

Since we assume that the fluid is inviscid and a potential flow, the flow is governed by an unsteady Bernoulli equation
\begin{equation}
    \frac{\partial\phi}{\partial t} + \frac{1}{2}u^2 - gz + \frac{p}{\rho} = \text{constant},
    \label{eq:bernoulli_generic}
\end{equation}
where $\rho$ is the fluid density, $u$ is the norm of the velocity, $p$ is the pressure, $g$ is the acceleration due to gravity and $z$ is the vertical coordinate below the initial fluid surface. This equation is constant in the entire fluid domain. Far from the crater, $u \to 0$, $\phi \to 0$ and the pressure is hydrostatic $p(z)=p_0+\rho g z$, where $p_0$ is the atmospheric pressure. This means that the constant is equal to $p_0/\rho$.

At the crater boundary, \textit{i.e.} at $r=R(\theta,t)$ (equation \ref{eq:crater_shape}), the Young-Laplace equation writes
\begin{equation}
    p(R)-p_0=\sigma C,
    \label{eq:young_laplace}
\end{equation}
where $C(\theta,t)$ is the mean local curvature of the interface and $\sigma$ the surface tension.
In cylindrical coordinates, the curvature writes
\begin{equation}
    C(\theta,t)=\frac{R^2+2\left(\frac{\partial R}{\partial\theta}\right)^2-R\frac{\partial^2R}{\partial\theta^2}}{\left[R^2+\left(\frac{\partial R}{\partial\theta}\right)^2\right]^{3/2}}.
\end{equation}
The Bernoulli equation at the crater boundary thus writes
\begin{equation}
    \left(\frac{\partial\phi}{\partial t}\right)_{r=R}+\frac{1}{2}u(R)^2-gR\cos\theta+\frac{\sigma}{\rho}C=0.
    \label{eq:bernoulli}
\end{equation}
We also use a kinematic boundary condition at the crater boundary
\begin{equation}
    \frac{\partial R}{\partial t}+\boldsymbol{u}\cdot\nabla R=\frac{\partial R}{\partial t}+u_\theta(R)\frac{1}{R}\frac{\partial R}{\partial \theta}=u_r(R).
    \label{eq:BC}
\end{equation}

Equations \ref{eq:bernoulli} and \ref{eq:BC} are made dimensionless using the scaling laws for the crater opening timescale $\tilde{t}_{max}$ (equation \ref{eq:energy_model_tmax}) and the maximum crater radius $\tilde{R}_{max}$ (equation \ref{eq:energy_model_Rmax}), which gives the partial differential equation system
\begin{equation}
    \left\{
    \begin{array}{r}
        \left(\frac{\partial\phi^*}{\partial t^*}\right)_{r^*=R^*}=-\frac{1}{2}u^*(R^*)^2+\frac{1}{4}\mathrm{B}\left(\frac{1}{2},\frac{5}{8}\right)^2 \xi R^*\cos\theta-\frac{1}{8}\sqrt{\frac{3}{2}}\mathrm{B}\left(\frac{1}{2},\frac{5}{8}\right)^2\frac{\xi}{\sqrt{\Phi}}\frac{\sqrt{Fr}}{We}C^* \\
        \frac{\partial R^*}{\partial t^*}=u^*_r(R^*)-u^*_\theta(R^*)\frac{1}{R^*}\frac{\partial R^*}{\partial \theta}
    \end{array}
    \right.,
    \label{eq:bernoulli_BC}
\end{equation}
where the star notation denotes quantities made dimensionless with $\tilde{R}_{max}$ and $\tilde{t}_{max}$, \textit{e.g.} $t^*=\tilde{t}/\tilde{t}_{max}$.

We solve this differential equation system (equations \ref{eq:bernoulli_BC}) by expanding the velocity potential (equation \ref{eq:potential_legendre}) up to degree $n_{max}=2$
\begin{equation}
    \phi^*(r^*,\theta,t^*)=\frac{\phi^*_0(t^*)}{{r^*}}+\frac{\phi^*_1(t^*)\cos\theta}{{r^*}^2}+\frac{\phi^*_2(t^*)\left(3\cos^2\theta-1\right)}{2{r^*}^3}.
    \label{eq:potential_legendre_truncated}
\end{equation}
The components of the velocity field then write (equation \ref{eq:velocity_legendre})
\begin{equation}
    \left\{
    \begin{array}{l}
        u^*_r(r^*,\theta,t^*)=-\frac{\phi^*_0(t^*)}{{r^*}^2}-\frac{2 \phi^*_1(t^*)\cos\theta}{{r^*}^3}-\frac{3\phi^*_2(t^*)\left(3\cos^2\theta-1\right)}{2{r^*}^4} \\
        u^*_\theta(r^*,\theta,t^*)=-\frac{\phi^*_1(t^*)\sin\theta}{{r^*}^3}-\frac{3\phi^*_2(t^*)\sin\theta\cos\theta}{{r^*}^4}
    \end{array}
    \right..
    \label{eq:velocity_legendre_truncated}
\end{equation}
We also expand the crater boundary position (equation \ref{eq:crater_shape}) up to degree $k_{max}=1$
\begin{equation}
    R^*(\theta,t^*)=R^*_0(t^*)+R^*_1(t^*)(2\cos\theta-1).
    \label{eq:crater_shape_truncated}
\end{equation}
Note that the crater position $R^*(\theta,t^*)$ is written as a sum of shifted Legendre polynomials, while the velocity potential $\phi^*(r^*,\theta,t^*)$ is a sum of standard Legendre polynomials.

We then project the differential equation system (equation \ref{eq:bernoulli_BC}) on a set of shifted Legendre polynomials $\bar{P}_m$ up to degree $m_{max}=2$ for the Bernoulli equation and degree $m_{max}=1$ for the kinematic boundary condition. The projection of a function $X$ writes
\begin{equation}
    \langle X,\bar{P}_m \rangle = (2m+1)\int_{0}^{\pi/2}X\bar{P}_m(\cos\theta)\sin\theta\mathrm{d}\theta.
    \label{eq:legendre_shifted_projection}
\end{equation}
We simplify the equations by expanding the Bernoulli equation and the kinematic boundary condition to the third and the fourth order in $R^*_1$. We obtain a system of five equations with five unknowns $\phi^*_0(t^*)$, $\phi^*_1(t^*)$, $\phi^*_2(t^*)$, $R^*_0(t^*)$ and $R^*_1(t^*)$ (equations \ref{eq:BC_0}-\ref{eq:bernoulli_2}).

The general equation system (equation \ref{eq:bernoulli_BC}) and its projection (equations \ref{eq:BC_0}-\ref{eq:bernoulli_2}) may be further simplified. The third term on the right-hand side of the Bernoulli equation (in equation \ref{eq:bernoulli_BC}) corresponds to surface tension effects associated with the curvature of the air-water interface. If this term is neglected, which corresponds to $\sqrt{Fr}/We \ll 1$, equation \ref{eq:bernoulli_BC} then simplifies as
\begin{equation}
    \left\{
    \begin{array}{r}
        \left(\frac{\partial\phi^*}{\partial t^*}\right)_{r^*=R^*}=-\frac{1}{2}u^*(R^*)^2+\frac{1}{4}\mathrm{B}\left(\frac{1}{2},\frac{5}{8}\right)^2 \xi R^*\cos\theta \\
        \frac{\partial R^*}{\partial t^*}=u^*_r(R^*)-u^*_\theta(R^*)\frac{1}{R^*}\frac{\partial R^*}{\partial \theta}
    \end{array}
    \right..
    \label{eq:bernoulli_BC_simplified}
\end{equation}
In our experiments, $\sqrt{Fr}/We$ is two to three times larger for case A ($1.0 \times 10^{-1}$) than for cases B, C and D ($4.9 \times 10^{-2}$, $3.9 \times 10^{-2}$ and $3.3 \times 10^{-2}$, respectively). This is consistent with the surface tension argument used to explain the difference between case A and the other cases (§\ref{sec:results}).
Since $\xi$ is independent of $Fr$ and $We$ in our experimental range \citep{lherm_2022}, this normalised equation system without surface tension is independent of the impact parameters and may be used to provide a predictive model.

The general and the simplified equation systems are solved numerically as initial value problems, using a differential equation solver. The solution thus depends on the choice of initial conditions.
On one hand, we can solve the equation systems separately for each experiment. The initial conditions are defined at $\tilde{t}=1$, which corresponds to an advection time of the impacting drop, and fitted on each experiment by using a joint least-square inversion of the five experimental coefficients over the entire time series.
On the other hand, we can solve the equation systems at the same time for all the experiments. The initial conditions are also defined at $\tilde{t}=1$ but fitted simultaneously on all the experiments using the joint least-square inversion over the entire time series. This method allows to define a unique set of initial conditions that may be used in a predictive model.
In both cases, the fitting procedure is motivated by the sensitivity of the model to the initial conditions used. A slight modification of the initial conditions may change significantly the time evolution of the coefficients. This sensitivity might be related to the exact impact conditions, including a possible variability in the contact dynamics with the surface of the pool and in the shape of the drop upon impact. Furthermore, the sensitivity to initial conditions might be amplified by the truncation of the crater shape and of the velocity potential expansion, which is probably insufficient to model properly the early evolution of the crater. This sensitivity is investigated in more detail in appendix \ref{app:initial_conditions_legendre}.

We now define two models using different systems of equations and definitions of initial conditions.
The first model, referred to as the general model, accounts for surface energy effects and uses the general equation system (equation \ref{eq:bernoulli_BC}) and initial conditions fitted on single experiments. This means that the number of sets of initial conditions is equal to the number of experiments. For example, the initial conditions of a given experiment in case B are $\phi^*_0(1)=-0.07 \pm 0.02$, $\phi^*_1(1)=-0.07 \pm 0.02$, $\phi^*_2(1)=0.009 \pm 0.003$, $R^*_0(1)=0.41 \pm 0.03$ and $R^*_1(1)=-0.28 \pm 0.02$. Uncertainties on the coefficients correspond to $1-\sigma$ standard deviations on the parameters in the least-square inversion. The initial conditions of all the experiments are presented in appendix \ref{app:initial_conditions_legendre}.
The second model, referred to as the simplified model, uses the simplified equation system, without surface tension and independent of the impact parameters (equation \ref{eq:bernoulli_BC_simplified}), as well as initial conditions fitted on all the experiments. The reference set of initial conditions is
\begin{equation}
    \left\{
    \begin{array}{lll}
        \phi^*_0(1)=-0.21 \pm 0.01,& \phi^*_1(1)=0.002 \pm 0.005,& \phi^*_2(1)=0.0004 \pm 0.0005,\\
        R^*_0(1)=0.29 \pm 0.02,& R^*_1(1)=-0.39 \pm 0.02.&
    \end{array}
    \right.
    \label{eq:reference_initial_conditions}
\end{equation}
Given the uncertainties, this set of initial conditions can be further simplified by using $\phi^*_1(1)=\phi^*_2(1)=0$, which corresponds to an initial velocity field given by $(u^*_r=-\phi^*_0(1)/{r^*}^2, u^*_\theta=0)$. The physical interpretation of these initial conditions should be investigated in the future. It probably involves the contact dynamics between the drop and the pool and the early evolution of the crater. Nonetheless, the simplified model is a predictive model, independent of the impact parameters, that can be used to predict the crater and velocity field evolution within the range of $Fr$ and $We$ covered by our experiments. However, we anticipate the model to show predictability limitations outside of this range, in particular at low $Fr$ and $We$ in the bubble entrapment region \citep[\textit{e.g.}][]{pumphrey_1990}, due to the neglected surface tension term and more generally to the relatively low degree of truncation used in our model.

\begin{figure}
    \centering
    \includegraphics[width=\linewidth]{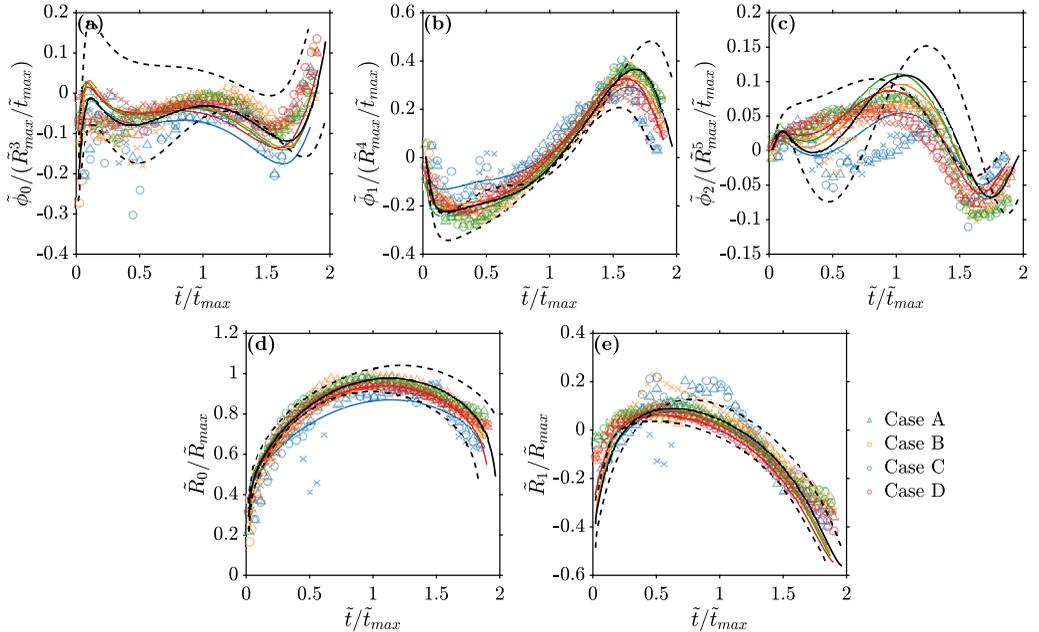}
    \caption{Coefficients $\phi^*_0$ (a), $\phi^*_1$ (b), $\phi^*_2$ (c), $R^*_0$ (d) and $R^*_1$ (e) as a function of time normalised by the opening timescale of the crater $\tilde{t}/\tilde{t}_{max}$, in the four cases. For each case, the different types of markers correspond to different experiments. The coloured solid lines give the solution of the general model, where the initial conditions are fitted on a single experiment of the corresponding case. The black solid lines give a solution to the simplified model, where the initial conditions are fitted simultaneously on all the experiments. The black dashed lines give a solution to the simplified model, where the initial conditions are modified by $\pm 25\%$ with respect to their reference value.}
    \label{fig:Fig11}
\end{figure}

Figure \ref{fig:Fig11} compares the experimental coefficients $\phi^*_0$ (a), $\phi^*_1$ (b) and $\phi^*_2$ (c) of the velocity potential and the experimental coefficients $R^*_0$ (d) and $R^*_1$ (e) of the crater shape with the coefficients obtained with the general (coloured solid lines) and the simplified (black solid lines) models.
We determine the experimental velocity potential coefficients from the experimental velocity field using a joint least-square inversion of the radial and the polar components (equation \ref{eq:velocity_legendre}). We also obtain the experimental crater shape coefficients by fitting the crater boundary position with the shifted Legendre polynomials expansion (equation \ref{eq:crater_shape}), using the method described in § \ref{sec:experiments_diagnostic_crater}.

The models capture well the evolution of the velocity potential (figure \ref{fig:Fig11}a-c) for all cases. In detail, the models are dominated by $\phi^*_1$ and are slightly less accurate when it comes to fit $\phi^*_2$, as expected since this corresponds to velocity fluctuations on smaller scales.
These results are consistent with the good agreement between the simplified model and the experimental velocity coefficients of figure \ref{fig:Fig10}, in particular regarding the slope of $u_{r,0}(r,t)$. Although $u_{r,0}(r,t)$ remains less steep than in the experiments, it decreases significantly faster than $1/r^2$.
The models also capture well the evolution of the crater shape (figure \ref{fig:Fig11}d-e). Note that $\tilde{R}_{max}$ slightly overestimates the experimental maximum crater radius, with maximum $R^*_0$ systematically smaller than 1. This can be explained by the neglected surface energy in the energy balance \citep{lherm_2022}.
In detail, $R^*_1$ is slightly underestimated and changes at a higher rate than experimental data when $\tilde{t}/\tilde{t}_{max} \lesssim 0.4$ and $\tilde{t}/\tilde{t}_{max} \gtrsim 1.7$. This corresponds respectively to the early opening of the crater and the end of crater collapse, including the formation of the central jet, where an expansion of $R$ to a higher degree (at least $k=2$) would be required to model the observed degree of deformation of the cavity (\textit{e.g.} figure \ref{fig:Fig4}c).

Note that the predictive model, using the simplified equation system (equation \ref{eq:bernoulli_BC_simplified}) with the reference set of initial conditions (equation \ref{eq:reference_initial_conditions}), is particularly in good agreement with the experimental data. The sensitivity of the simplified model to the initial conditions is illustrated with two solutions where the initial conditions have been modified by $\pm 25\%$ with respect to their reference value (figure \ref{fig:Fig11}, black dashed lines).

Although case A is slightly different from cases B, C and D due to surface tension effects (see §\ref{sec:results}), the models capture properly the general cratering dynamics. In detail, $\phi^*_2$ and $R^*_1$ are significantly underestimated when $0.5 \lesssim \tilde{t}/\tilde{t}_{max} \lesssim 1.4$, as expected since the models do not account for the capillary wave propagation responsible for this cavity deformation.

\begin{figure}
    \centering
    \includegraphics[width=\linewidth]{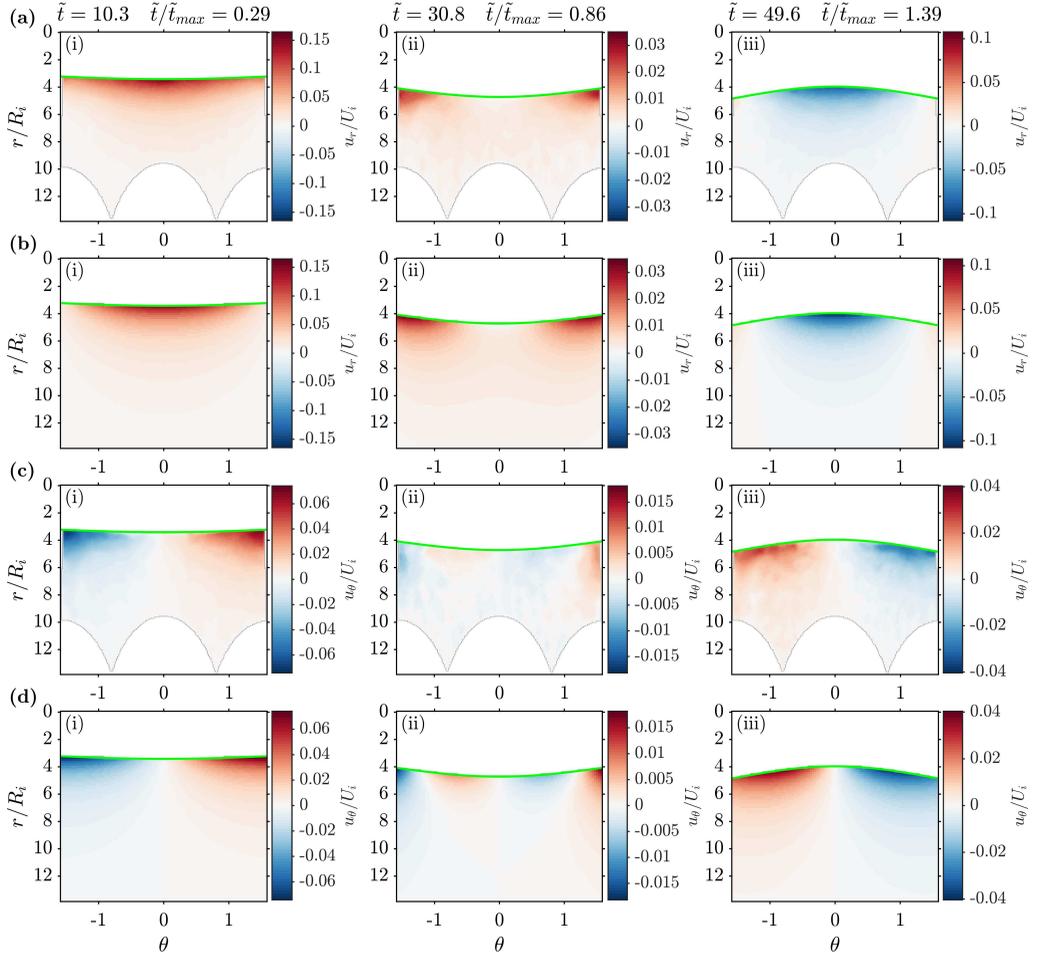}
    \caption{Radial (a, b) and polar (c, d) component of the velocity field from experimental data (a, c) and from the simplified model (b, d), in case B. The snapshots correspond to times when the crater is opening (i), when the crater is almost at its maximum size (ii) and when the crater is closing (iii). The solid green lines correspond to the experimental crater boundary.}
    \label{fig:Fig12}
\end{figure}

Figure \ref{fig:Fig12} compares snapshots of the radial (a-b) and polar components (c-d) of the experimental velocity field (a-c) with the components calculated from the predictive simplified model (b-d), in case B. The comparison is conducted at different times during the opening stage (i), just before the crater reaches its maximum size (ii), and during the closing stage (iii). This illustrates that the velocity fields from the simplified model and the experiment are very similar during all stages of the cratering process. The differences observed are mainly in the magnitude of the velocity, in particular close to the crater and the initial water surface ($\theta = \pm \pi/2$). Similar results are obtained in the other cases.
The good agreement between the experimental velocity field and the simplified model shows that the truncation used in the model (degree $k=1$ in shifted Legendre polynomials for $R^*$ and degree $n=2$ in Legendre polynomials for $\phi^*$) is sufficient to accurately capture the flow dynamics.

\begin{figure}
    \centering
    \includegraphics[width=\linewidth]{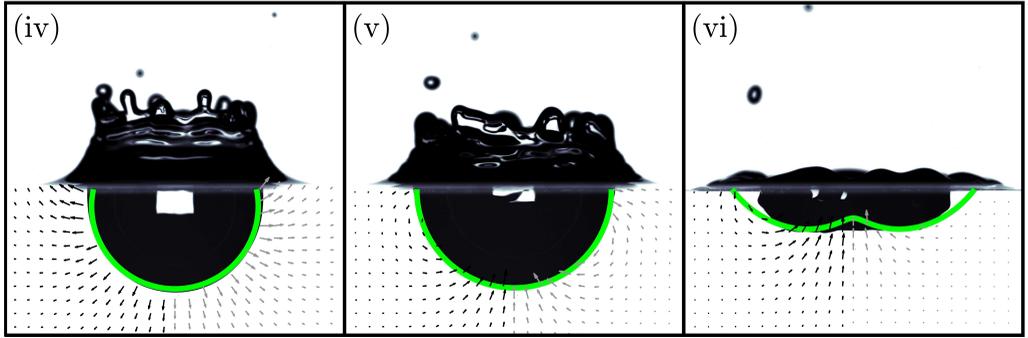}
    \caption{Time evolution of the crater shape obtained in a backlight experiment similar to case B ($Fr=442$). The solid green lines correspond to the crater boundary obtained from the simplified model. The black arrows correspond to the experimental velocity field, normalised by its maximum value in each snapshot. The grey arrows correspond to the velocity field obtained from the simplified model, normalised by its maximum value in each snapshot.}
    \label{fig:Fig13}
\end{figure}

Figure \ref{fig:Fig13} compares the crater shape obtained in a backlight experiment similar to case B ($Fr=442$) with the crater boundary position calculated from the predictive simplified model. 
The crater shape is well captured by the model, consistently with the results of figure \ref{fig:Fig11}d-e. 
In detail, at the very beginning of the crater opening stage (figure \ref{fig:Fig13}, i), the model overestimates the width of the crater and does not capture accurately the flat-bottomed shape of the cavity. 
During the crater opening stage and the beginning of the crater collapse stage (figure \ref{fig:Fig13}, iii-v), the model slightly underestimates the crater depth and width, consistently with the coefficients of figure \ref{fig:Fig11}d-e. 
Finally, when the crater collapses (figure \ref{fig:Fig13}, vi), the model shows the central jet initiation, although it visibly lacks higher degrees to account for the vertical walls of the cavity.
Figure \ref{fig:Fig13} also compares the experimental velocity field obtained in case B with the velocity field obtained from the simplified model. The comparison shows a good agreement between the two, which is consistent with the analysis of figure \ref{fig:Fig12}.

\section{Conclusion}
\label{sec:conclusion}

In this paper, we analyse quantitatively the velocity field around the crater produced by the impact of a liquid drop onto a deep liquid pool. Using new high-resolution PIV measurements, we obtain simultaneously the evolution of the velocity field around the cavity and the crater shape.
We found that the shape of the cavity and the velocity field seem to be independent of $Fr$ and $We$ at a given $t/t_{max}$, when these two dimensionless numbers are large enough (cases B, C and D).
The velocity field is dominated by the degrees 0 and 1 in terms of shifted Legendre polynomials, with the degree 0 of the radial component $u_{r,0}(r,t)$ decreasing faster than $1/r^2$.
Furthermore, the radial component of the velocity field is dominated by the degree 1 in terms of standard Legendre polynomials. This is not inconsistent with the growth of the crater because the degree 1 of the radial component has a non-zero average over a hemisphere.
The experiments also show significant contributions from the degree 2, in particular when the crater is strongly deformed. This is possibly related to the non-hemispherical shape (degree 1) of the cavity. 
We also found that the velocity field does not vanish when the crater reaches its maximum size.

In the previous velocity models \citep{engel_1967,maxwell_1977,leng_2001,bisighini_2010}, strong constraints were imposed on the crater shape and/or on the velocity field. They were unable to explain the properties observed in our experimental measurements, in particular the radial dependency of the radial component of the velocity field and the evolution of the shape of the cavity.
We thus developed a semi-analytical model based on a Legendre polynomials expansion of an unsteady Bernoulli equation, coupled with a kinematic boundary condition at the crater boundary. Assuming that the surface tension term involved in the Bernoulli equation is negligible, we define a simplified model, independent of the impact parameters, that can predict the evolution of the crater shape and of the velocity field within the range of $Fr$ and $We$ numbers covered in our experiments. Although the model is sensitive to the initial conditions, it remains predictive by using a unique set of fitted initial conditions.
In particular, the model can capture the initiation of the central jet. However, one intrinsic limitation of the model is that it assumes the cavity radius to be a bijective function of $\theta$. While this assumption is true during the opening stage and part of the crater closing stage, including the central jet initiation, it eventually fails when the central jet reaches a critical height, since the air/water interface can be crossed twice at a given $\theta$. The model can therefore not be used to describe the full growth of the central jet.

% \backsection[Supplementary data]{\label{SupMat}Supplementary movies are available at \\https://doi.org/10.1017/jfm.2019...}

\backsection[Acknowledgements]{We thank M. Moulin for his help with the design and construction of the experimental apparatus. We thank three anonymous reviewers for their valuable comments which significantly improved the manuscript.}

\backsection[Funding]{This work was supported by the European Research Council (ERC) under the European Unions Horizon 2020 research and innovation programme (grant number 716429). ISTerre is part of Labex OSUG@2020 (ANR10 LABX56). Partial funding for this research was provided by the Center for Matter at Atomic Pressure (CMAP), a National Science Foundation (NSF) Physics Frontier Center, under award PHY-2020249. Any opinions, findings, conclusions or recommendations expressed in this material are those of the author(s) and do not necessarily reflect those of the National Science Foundation.}

\backsection[Declaration of interests]{The authors report no conflict of interest.}

%\backsection[Data availability statement]{The data that support the findings of this study are openly available in [repository name] at http://doi.org/[doi], reference number [reference number].}

\backsection[Author ORCID]{V. Lherm, https://orcid.org/0000-0001-5814-0637; R. Deguen, https://orcid.org/0000-0002-6883-0655}

\backsection[Author contributions]{V.L. and R.D. designed the study, derived the model and contributed to analysing data and reaching conclusions. V.L. conducted the experiments. V.L. and R.D wrote the paper.}

\appendix

\section{Equations of the Legendre polynomials model}
\label{app:equations_legendre_model}

The Legendre polynomials model equations correspond to the projection of equation \ref{eq:bernoulli_BC} up to degree $m_{max}=1$ for the kinematic boundary condition and up to degree $m_{max}=2$ for the Bernoulli equation. The projected boundary conditions and Bernoulli equations are respectively expanded to the fourth and the third order in $R^*_1$. The boundary condition then writes
\begin{eqnarray}
    \label{eq:BC_0}
    \nonumber \dot{R}^*_0&=&\frac{1}{6 {R^*_0}^8}\left\{-6 {\phi^*_0} {R^*_0}^2 \left({R^*_0}^2 {R^*_1}^2+{R^*_0}^4+{R^*_1}^4\right)-2 {\phi^*_1} {R^*_0} \left({R^*_0}^3 {R^*_1}+10 {R^*_0}^2 {R^*_1}^2+6 {R^*_0} {R^*_1}^3\right.\right.\\
    \nonumber &&\left.\left.+3 {R^*_0}^4+21 {R^*_1}^4\right)+3 {\phi^*_2} {R^*_1} (3 {R^*_0}-{R^*_1}) \left({R^*_0}^2+3 {R^*_1}^2\right)\right\}\\
    &&+O({R^*_1}^5),\\
    \label{eq:BC_1}
    \nonumber \dot{R}^*_1&=&\frac{1}{140 {R^*_0}^8}\left\{14 {R^*_0}^3 {R^*_1} [20 {R^*_0} ({\phi^*_0} {R^*_0}+2 {\phi^*_1})+9 {\phi^*_2}]+6 {R^*_0} {R^*_1}^3 [56 {R^*_0} ({\phi^*_0} {R^*_0}+5 {\phi^*_1})+75 {\phi^*_2}]\right.\\
    \nonumber &&\left.+84 {R^*_0}^2 {R^*_1}^2 (2 {\phi^*_1} {R^*_0}-15 {\phi^*_2})+15 {R^*_1}^4 (68 {\phi^*_1} {R^*_0}-189 {\phi^*_2})-35 {R^*_0}^4 (4 {\phi^*_1} {R^*_0}+9 {\phi^*_2})\right\}\\
    &&+O({R^*_1}^5),
\end{eqnarray}
and the Bernoulli equation writes
\begin{eqnarray}
    \label{eq:bernoulli_0}
    \nonumber 0&=&\frac{1}{1680 {R^*_0}^{11}}\left\{28 {R^*_0}^2 \left[{R^*_0}^3 \left(20 {R^*_0}^3 \dot{\phi}^*_0 \left(3 {R^*_0}^2+{R^*_1}^2\right)+10 {\phi^*_0}^2 \left(3 {R^*_0}^2+10 {R^*_1}^2\right)\right.\right.\right.\\
    \nonumber &&\left.\left.\left.+2 {R^*_0} \dot{\phi}^*_1 \left(-10 {R^*_0}^2 {R^*_1}+15 {R^*_0} {R^*_1}^2+15 {R^*_0}^3-12 {R^*_1}^3\right)-3 {R^*_1} \dot{\phi}^*_2 \left(-4 {R^*_0} {R^*_1}+15 {R^*_0}^2\right.\right.\right.\right.\\
    \nonumber &&\left.\left.\left.\left.+30 {R^*_1}^2\right)\right)+20 {\phi^*_0} {\phi^*_1} {R^*_0} \left(-5 {R^*_0}^2 {R^*_1}+15 {R^*_0} {R^*_1}^2+3 {R^*_0}^3-21 {R^*_1}^3\right)+6 {\phi^*_1}^2 \left(-15 {R^*_0}^2 {R^*_1}\right.\right.\right.\\
    \nonumber &&\left.\left.\left.+77 {R^*_0} {R^*_1}^2+10 {R^*_0}^3-84 {R^*_1}^3\right)\right]+168 {\phi^*_2} {R^*_0} \left[-9 {R^*_0}^2 {R^*_1} (5 {\phi^*_0} {R^*_0}+7 {\phi^*_1})+7 {R^*_0} {R^*_1}^2 (3 {\phi^*_0} {R^*_0}\right.\right.\\
    \nonumber &&\left.\left.+28 {\phi^*_1})-36 {R^*_1}^3 (7 {\phi^*_0} {R^*_0}+13 {\phi^*_1})+15 {\phi^*_1} {R^*_0}^3\right]+72 {\phi^*_2}^2 \left[-70 {R^*_0}^2 {R^*_1}+558 {R^*_0} {R^*_1}^2\right.\right.\\
    \nonumber &&\left.\left.+35 {R^*_0}^3-720 {R^*_1}^3\right]-\frac{35}{\sqrt{\Phi} We}\mathrm{B}\left(1/2,5/8\right)^2 \xi {R^*_0}^6 \left[-3 \sqrt{6} \sqrt{Fr} {R^*_0}^2 {R^*_1}^2\right.\right.\\
    \nonumber &&\left.\left.+3 \sqrt{6} \sqrt{Fr} {R^*_0} {R^*_1} \left({R^*_1}^2-{R^*_0}^2\right)-3 \sqrt{6} \sqrt{Fr} {R^*_0}^4+2 \sqrt{\Phi} We {R^*_0}^5 {R^*_1}+6 \sqrt{\Phi} We {R^*_0}^6\right]\right\}\\
    &&+O({R^*_1}^4),
\end{eqnarray}
\begin{eqnarray}
    \label{eq:bernoulli_1}
    \nonumber 0&=&\frac{1}{280 {R^*_0}^{11}}\left\{-14 {R^*_0}^2 \left[4 {R^*_0}^2 {R^*_1} \left({R^*_0}^3 \dot{\phi}^*_0 \left(5 {R^*_0}^2+3 {R^*_1}^2\right)+10 {\phi^*_0}^2 \left({R^*_0}^2+3 {R^*_1}^2\right)\right)\right.\right.\\
    \nonumber &&\left.\left.-20 {\phi^*_0} {\phi^*_1} {R^*_0} \left(-5 {R^*_0}^2 {R^*_1}+9 {R^*_0} {R^*_1}^2+{R^*_0}^3-21 {R^*_1}^3\right)+2 {R^*_0}^4 \dot{\phi}^*_1 \left(10 {R^*_0}^2 {R^*_1}-9 {R^*_0} {R^*_1}^2\right.\right.\right.\\
    \nonumber &&\left.\left.\left.-5 {R^*_0}^3+12 {R^*_1}^3\right)+3 {\phi^*_1}^2 \left(44 {R^*_0}^2 {R^*_1}-63 {R^*_0} {R^*_1}^2-5 {R^*_0}^3+256 {R^*_1}^3\right)\right]\right.\\
    \nonumber &&\left.+42 {\phi^*_2} {R^*_0} \left[3 {\phi^*_0} {R^*_0} \left(-4 {R^*_0}^2 {R^*_1}+63 {R^*_0} {R^*_1}^2+5 {R^*_0}^3-32 {R^*_1}^3\right)+2 {\phi^*_1} \left(-49 {R^*_0}^2 {R^*_1}\right.\right.\right.\\
    \nonumber &&\left.\left.\left.+156 {R^*_0} {R^*_1}^2+9 {R^*_0}^3-396 {R^*_1}^3\right)\right]+6 {R^*_0}^5 \dot{\phi}^*_2 \left[-14 {R^*_0}^2 {R^*_1}+126 {R^*_0} {R^*_1}^2+35 {R^*_0}^3-40 {R^*_1}^3\right]\right.\\
    \nonumber &&\left.+9 {\phi^*_2}^2 \left[-496 {R^*_0}^2 {R^*_1}+864 {R^*_0} {R^*_1}^2+35 {R^*_0}^3-4960 {R^*_1}^3\right]\right.\\
    \nonumber &&\left.-\frac{1120 \pi}{\sqrt{\Phi} We }\frac{\Gamma \left(5/8\right)^2}{\Gamma \left(1/8\right)^2} \xi {R^*_0}^6 ({R^*_0}+{R^*_1}) \left[3 \sqrt{6} \sqrt{Fr} {R^*_0} {R^*_1}^2+2 \sqrt{\Phi} We {R^*_0}^5\right]\right\}\\
    &&+O({R^*_1}^4),
\end{eqnarray}
\begin{eqnarray}
    \label{eq:bernoulli_2}
    \nonumber 0&=&\frac{1}{336 {R^*_0}^{11}}\left\{{R^*_0}^2 \left[{R^*_0}^3 \left(4 \left(56 {R^*_1}^2 \left({R^*_0}^3 \dot{\phi}^*_0+5 {\phi^*_0}^2\right)-4 {R^*_0} {R^*_1} \dot{\phi}^*_1 \left(-21 {R^*_0} {R^*_1}+14 {R^*_0}^2\right.\right.\right.\right.\right.\\
    \nonumber &&\left.\left.\left.\left.\left.+24 {R^*_1}^2\right)+3 \dot{\phi}^*_2 \left(-42 {R^*_0}^2 {R^*_1}+22 {R^*_0} {R^*_1}^2+7 {R^*_0}^3-120 {R^*_1}^3\right)\right)\right.\right.\right.\\
    \nonumber &&\left.\left.\left.-\frac{7}{\sqrt{\Phi} We}\mathrm{B}\left(1/2,5/8\right)^2 \xi {R^*_1} \left(3 \sqrt{6} \sqrt{Fr} {R^*_0}^3 {R^*_1}-21 \sqrt{6} \sqrt{Fr} {R^*_0}^2 {R^*_1}^2\right.\right.\right.\right.\\
    \nonumber &&\left.\left.\left.\left.+4 \sqrt{\Phi} We {R^*_0}^6\right)\right)-1120 {\phi^*_0} {\phi^*_1} {R^*_0} {R^*_1} \left(-3 {R^*_0} {R^*_1}+{R^*_0}^2+6 {R^*_1}^2\right)+84 {\phi^*_1}^2 \left(-12 {R^*_0}^2 {R^*_1}\right.\right.\right.\\
    \nonumber &&\left.\left.\left.+67 {R^*_0} {R^*_1}^2+{R^*_0}^3-96 {R^*_1}^3\right)\right]+84 {\phi^*_2} {R^*_0} \left[3 {\phi^*_0} {R^*_0} \left(-12 {R^*_0}^2 {R^*_1}+11 {R^*_0} {R^*_1}^2+{R^*_0}^3\right.\right.\right.\\
    \nonumber &&\left.\left.-96 {R^*_1}^3\right)+2 {\phi^*_1} ({R^*_0}-6 {R^*_1}) \left(-9 {R^*_0} {R^*_1}+3 {R^*_0}^2+46 {R^*_1}^2\right)\right]+18 {\phi^*_2}^2 \left[-148 {R^*_0}^2 {R^*_1}\right.\\
    \nonumber &&\left.\left.+1116 {R^*_0} {R^*_1}^2+23 {R^*_0}^3-1860 {R^*_1}^3\right]\right\}\\
    &&+O({R^*_1}^4).
\end{eqnarray}
The simplified version of the equation system (equation \ref{eq:bernoulli_BC_simplified}) can be obtained by using $\sqrt{Fr}/We=0$ in equations \ref{eq:BC_0}-\ref{eq:bernoulli_2}.

\section{Initial conditions of the Legendre polynomials model}
\label{app:initial_conditions_legendre}

\begin{figure}
    \centering
    \includegraphics[width=\linewidth]{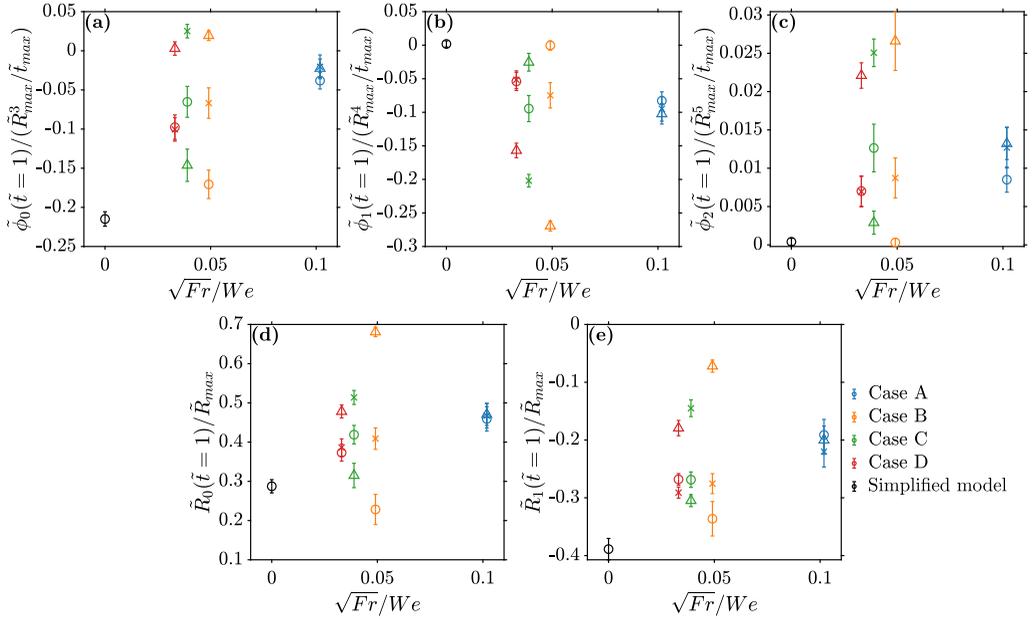}
    \caption{Initial conditions (at $\tilde{t}=1$) of the Legendre polynomials models, for the coefficients $\phi^*_0$ (a), $\phi^*_1$ (b), $\phi^*_2$ (c), $R^*_0$ (d) and $R^*_1$ (e), obtained using a joint least-square inversion of the experimental coefficients over the entire time series. The coloured markers correspond to the general model, where initial conditions are obtained by fitting experiments individually. The black circles correspond to the simplified model, where initial conditions are obtained by fitting all the experiments simultaneously. Uncertainties correspond to $1-\sigma$ standard deviations on the parameters in the least-square inversion.}
    \label{fig:Fig14}
\end{figure}

Figure \ref{fig:Fig14} shows the initial conditions of the general model, obtained by fitting individually the experiments, and of the simplified model, obtained by fitting all the experiments simultaneously. They are both defined at $\tilde{t}=1$ and use a joint least-square inversion of the experimental coefficients over the entire time series. Uncertainties on the initial conditions correspond to $1-\sigma$ standard deviations on the parameters in the least-square inversion.

At low $\sqrt{Fr}/We$, corresponding to high $Fr$ and $We$ numbers (cases B, C, D), the dispersion of the initial conditions is larger than the uncertainties associated with the least-square inversion, whereas at higher $\sqrt{Fr}/We$, corresponding to moderate $Fr$ and $We$ numbers (case A), the initial conditions are clustered within the inversion uncertainties. This dispersion at higher $Fr$ and $We$ suggests a higher variability of the crater shape and of the velocity field upon impact. This might be related to a greater sensitivity to the exact impact conditions, possibly including variability in the contact dynamics with the surface of the pool and in the shape of the drop upon impact. Furthermore, we do not find any secondary dependency on $Fr$ or $We$. Finally, the initial conditions of the simplified model, obtained by fitting all the experiments simultaneously, are similar to the initial conditions obtained by fitting individually the experiments.

\begin{figure}
    \centering
    \includegraphics[width=\linewidth]{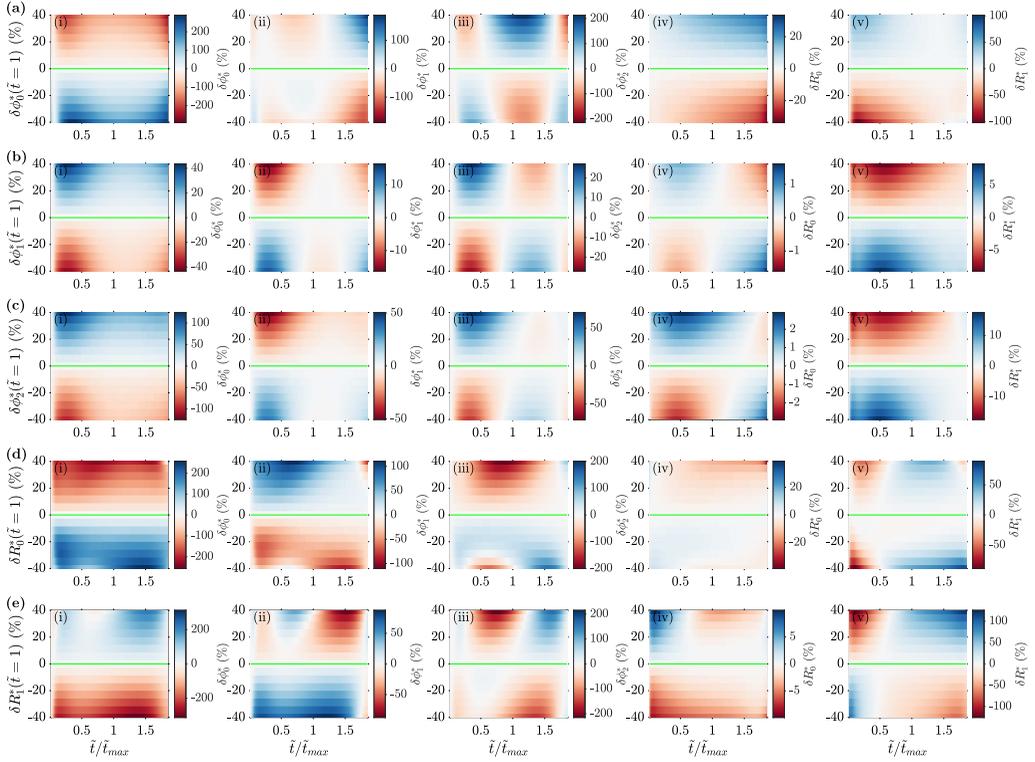}
    \caption{Sensitivity of the simplified model to individual changes in the initial conditions. Each row corresponds to the relative change $\delta X(1)=[X(1)-X_\mathrm{ref}(1)]/X_\mathrm{ref}(1)$ (in $\%$) of a single initial condition $X(1)=\{\phi^*_0(1), \phi^*_1(1), \phi^*_2(1), R^*_0(1), R^*_1(1)\}$ with respect to the reference value $X_\mathrm{ref}(1)$, defined in equation \ref{eq:reference_initial_conditions}. Each column corresponds to the relative change $\delta X=(X-X_\mathrm{ref})/\mathrm{RMS}(X_\mathrm{ref})$ (in $\%$) of the coefficient $X$ over the entire time series, for a given initial condition modification. The relative change $\delta X$ is defined as the absolute change in $X$, $X-X_\mathrm{ref}$, scaled by the root mean square of the simplified model $\mathrm{RMS}(X_\mathrm{ref})$.}
    \label{fig:Fig15}
\end{figure}

The relatively large dispersion observed for a given case (except for case A) indicates that the model is sensitive to the initial conditions. For example, a change in all the initial conditions by $\pm 25\%$ gives a significantly modified evolution of the coefficients over time (figure \ref{fig:Fig11}, black dashed lines).
In order to further investigate this initial condition sensitivity, we conducted a quantitative test on the simplified model. Figure \ref{fig:Fig15} shows the relative change of the model coefficients with respect to the simplified model, as a result of an individual modification of a single initial condition from the reference value defined in equation \ref{eq:reference_initial_conditions}. The relative change $\delta X$ is defined as the absolute change in $X=\{\phi^*_0, \phi^*_1, \phi^*_2, R^*_0, R^*_1\}$, $X-X_\mathrm{ref}$, scaled by the root mean square of the simplified model $\mathrm{RMS}(X_\mathrm{ref})$. We choose to scale the absolute change by the root mean square of the simplified model to ensure a non-diverging value of the relative change when $X_\mathrm{ref} \to 0$.
Note that this sensitivity test only investigates the role of independent parameter modifications. Coupled modifications of the initial conditions (as in figure \ref{fig:Fig11}, black dashed lines) might amplify significantly the changes in the evolution of the coefficients.

Within the range of parameter modifications (by $\pm 40\%$), the coefficients are generally more influenced by modifications of the initial conditions of the crater shape, \textit{i.e.} $R^*_0(1)$ (figure \ref{fig:Fig15}d) and $R^*_1(1)$ (figure \ref{fig:Fig15}e).
Besides, the coefficient $R^*_0$ is the least modified with a maximum change of $\sim 30\%$ (figure \ref{fig:Fig15}iv), while $\phi^*_0$, $\phi^*_1$, $\phi^*_2$ and $R^*_1$ reach respectively $\sim 300\%$ (figure \ref{fig:Fig15}i), $\sim 150\%$ (figure \ref{fig:Fig15}ii), $\sim 200\%$ (figure \ref{fig:Fig15}iii) and $\sim 100\%$ (figure \ref{fig:Fig15}v). Finally, the change in the coefficients over time is not homogeneous. For example, $\phi^*_0$ is changed relatively uniformly over time (in magnitude), independently of the modified initial condition, while $\phi^*_2$ is changed much more heterogeneously and depends on the modified initial condition.

\bibliographystyle{jfm}
\bibliography{biblio}
%Use of the above commands will create a bibliography using the .bib file. Shown below is a bibliography built from individual items.

%% End of file `jfm2esam.bib'.

\end{document}